\documentclass{article}

\usepackage{PRIMEarxiv}

\usepackage[utf8]{inputenc} 
\usepackage[T1]{fontenc}    
\usepackage{hyperref}       
\usepackage{url}            
\usepackage{booktabs}       
\usepackage{amsfonts}       
\usepackage{nicefrac}       
\usepackage{microtype}      
\usepackage{lipsum}
\usepackage{fancyhdr}       
\usepackage{graphicx}       
\graphicspath{{media/}}     

\title{Nuclear Arms Control Verification \\
and Lessons for AI Treaties
}

\author{
  Mauricio Baker \\
  Stanford University \\
  maubaker@stanford.edu \\
}

\begin{document}
\maketitle

\begin{abstract}
Security risks from AI have motivated calls for international agreements that guardrail the technology. However, even if states could agree on what rules to set on AI, the problem of verifying compliance might make these agreements infeasible. To help clarify the difficulty of verifying agreements on AI—and identify actions that might reduce this difficulty—this report examines the case study of verification in nuclear arms control. We review the implementation, track records, and politics of verification across three types of nuclear arms control agreements. Then, we consider implications for the case of AI, especially AI development that relies on thousands of highly specialized chips. In this context, the case study suggests that, with certain preparations, the foreseeable challenges of verification would be reduced to levels that were successfully managed in nuclear arms control. To avoid even worse challenges, substantial preparations are needed: (1) developing privacy-preserving, secure, and acceptably priced methods for verifying the compliance of hardware, given inspection access; and (2) building an initial, incomplete verification system, with authorities and precedents that allow its gaps to be quickly closed if and when the political will arises.
\end{abstract}

\section*{Executive Summary} \label{executivesummary}

\subsection*{Why nuclear arms control verification may matter for AI} \label{whynucleararmscontrolverificationmaymatterforai}

Along with its benefits, AI poses significant security threats. Current or upcoming AI applications risk escalating conflicts, boosting disinformation and terrorist activities, and causing large-scale accidents. Some researchers have proposed that states should address these risks by voluntarily joining international agreements on AI. However, even if states could agree on how to regulate AI, states may struggle to verify compliance. Verification is far from the only issue that could make coordinated rules on AI infeasible or undesirable, but it could be a particularly thorny issue. \textbf{To help clarify the challenge of verifying AI treaties, this report examines a relevant case study: nuclear arms control.}

The challenges of verifying AI treaties will plausibly have \textbf{significant similarities} to challenges in nuclear arms control verification. These include restricting both government and corporate activity, protecting sensitive information, countering well-resourced adversaries, and verifying nationwide inventories of dual-use items. On the other hand, states are likely less willing to accept costly verification for AI treaties, as AI security currently tends to be a relatively low priority. That difference, though, might diminish as AI security risks become more pressing. Given the plausible similarities, we proceed with the case study.

\subsection*{How monitoring \& verification (M\&V) has been implemented in nuclear arms control} \label{howmonitoringverificationmvhasbeenimplementedinnucleararmscontrol}

First, we review M\&V implementation across three types of nuclear arms control treaties: nonproliferation treaties; U.S.-U.S.S.R./Russia arms limitation treaties; and nuclear test bans.\footnote{This review mainly draws from a textbook about nonproliferation M\&V and the text of relevant treaties. Nonproliferation treaties are: the Non-Proliferation Treaty (NPT) and five Nuclear-Weapon-Free-Zone Treaties (i.e. Tlatelolco, Rarotonga, Bangkok, Pelindaba, and Semipalatinsk). (There were also agreements that were not formal treaties, mainly the U.S.-D.P.R.K. Agreed Framework and the JCPOA.) Nuclear arms limitation treaties are the: ABM Treaty, SALT I provisional agreement, INF Treaty, START, SORT, and New START. Test ban treaties are the: Partial Nuclear Test Ban Treaty, Threshold Test Ban Treaty, Peaceful Nuclear Explosions Treaty, and Comprehensive Nuclear-Test-Ban Treaty. (The latter would authorize on-site inspections to investigate events detected by remote sensor stations, but it is not on track to enter into force.)}

\textbf{Nuclear nonproliferation treaties} commit all but 10 states to not acquiring nuclear weapons\footnote{Of these ten states, five are parties to the NPT as nuclear weapon states (a status which permits them to keep their nuclear weapons), and another five are not parties to any nuclear nonproliferation treaty.}. They are mainly verified as follows:

\begin{itemize}
	\item States are required to declare (i.e. self-report) the amount and locations of all the nuclear material in their territory. (They may have nuclear material for nuclear energy.)
	\item To verify that nuclear materials and equipment in \emph{declared} nuclear facilities are not being used for weapons, international inspectors regularly verify declared accounts of nuclear materials. This involves much on-site measurement, verification of facility layouts, and surveillance at facilities. Under some agreements\footnote{These are Additional Protocols.}, certain non-nuclear facilities (e.g. adjacent facilities) are also declared and inspected.

	\item To verify that a state does not have \emph{undeclared} (i.e. secret) nuclear facilities, international inspectors receive voluntary tips from national intelligence agencies (which have e.g. spies and satellites), and they analyze their own limited information. Then, after identifying a suspect site, they investigate it (though this happens much more frequently under one type of agreement\footnote{That is, Additional Protocols, under which there is a much lower political bar to conducting such inspections.}).
\end{itemize}
\textbf{U.S.-U.S.S.R./Russia arms limitation treaties} limited the number and/or types of these states’ ready-to-use nuclear weapons. They were mainly verified as follows:

\begin{itemize}
	\item Under some treaties, the U.S. and U.S.S.R./Russia just used satellites (and presumably spies) to track each other’s ready-to-use nuclear weapons.
	\item Under other treaties, these states also self-reported the number, types, and locations of all their treaty-limited nuclear weapons. Then, each state regularly verified the other’s self-reports through satellites and on-site inspections (which mostly involved simple measurement), along with using radar to track missile test-flights.

\end{itemize}
\textbf{Nuclear weapon test ban treaties} are mainly verified as follows:

\begin{itemize}
	\item Sensors, including an international network of ~300 sensor stations, detect the acoustic waves and air particles that nuclear tests make, and analysts infer their source location.

\end{itemize}
\subsection*{Track records of M\&V in nuclear arms control} \label{trackrecordsofmvinnucleararmscontrol}

\textbf{The strongest widely implemented M\&V system in each of the above three categories of treaties}\footnote{These are: for nonproliferation, the IAEA system under Comprehensive Safeguards Agreements with Additional Protocols; for bilateral treaties, the INF Treaty, START, and New START systems; and for test bans, the International Monitoring System.}\textbf{ has had zero known major failures.} This was in the context of (up to) a few known attempts at serious violations (e.g. developing a banned type of missile), all of which the system detected. However, \textbf{weaker predecessors of these systems had some failures} at detecting major violations. Most notably, a system\footnote{That is, the IAEA’s Comprehensive Safeguards Agreements, especially before the 90s} for verifying nonproliferation was designed with a focus on declared nuclear facilities, and it repeatedly failed to detect undeclared ones (though the associated weapon programs were all discovered or abandoned before they developed nuclear weapons).\footnote{The review of track records is mainly based on reports from the NTI and IAEA.}

\subsection*{Politics of M\&V in nuclear arms control} \label{politicsofmvinnucleararmscontrol}

M\&V negotiation records and outcomes, along with negotiators’ incentives, suggest M\&V negotiations involved pressures to: ensure effectiveness, protect state and commercial secrets, limit disruptions and financial costs, avoid security threats, preserve national industrial competitiveness, keep compliance feasible, observe privacy rights, avoid passing new legislation, appease idiosyncratic stakeholders, limit partiality, and respect national sovereignty.\footnote{This takeaway mainly draws from: reports on negotiations, U.S. chief negotiator Gottemoeller’s account, and negotiation outcomes; these sources leave out many details of negotiations, so this section likely misses important aspects of M\&V politics.}

Separately, some of the failures mentioned earlier were influential. In particular, Iraq nearly developed nuclear weapons in spite of inspections, and this drove policymakers to greatly strengthen nonproliferation M\&V.

\subsection*{Lessons for the case of AI} \label{lessonsforthecaseofai}

First, let us narrow the scope of our analysis. While many AI activities carry risks, we will consider lessons for verifying rules\footnote{More concretely, here are two example rules states may want to uphold (without taking a stance on their wisdom): (1) The training of a large machine learning model should stop if the model’s offensive cyber capabilities reach a certain level, as measured by some objective benchmark. (2) Large machine learning models used in autonomous weapon systems must be trained with certain measures (e.g. adversarial robustness training) to reduce unintended behavior.} on one increasingly important AI activity: training machine learning models with industrial-scale, specialized computer chips. Additionally, we will consider one approach to verifying these rules: using mechanisms built onto cutting-edge chips. Ongoing research by e.g. Shavit  \cite{shavit2023} suggests such verification may be technically feasible, even while preserving privacy and efficiency (though that requires further R\&D).

In this context, our case study suggests that, \textbf{with certain preparations, the main foreseeable challenges of hardware-based AI treaty verification would be ones that were manageable in nuclear arms control.} To see this, we will consider each of the main challenges that hardware-based verification appears likely to face, based on nuclear M\&V politics. We will see that, for each of these challenges, substantial preparations would reduce it to a difficulty that was manageable\footnote{That is, some nuclear arms control M\&V system faced a similar or greater difficulty, yet it was adopted and successful.} in the nuclear case.

As the first challenge we consider, AI chip users may oppose verification due to concerns that it would expose sensitive data and software to spies and saboteurs. \textbf{Certain preparations would reduce these secrecy and security concerns to concerns that were manageable} in nuclear arms control.

\begin{itemize}
	\item States may need to disclose data centers’ locations, which could be sensitive. However, similar or worse concerns were manageable in the nuclear case; nearly all states agreed to disclose the locations of their nuclear energy facilities, and the U.S. and U.S.S.R. even agreed to share the locations of their nuclear weapon bases.
	\item If stakeholders develop privacy-preserving and secure methods for inspecting AI chips, then AI chip users’ secrecy and security concerns could be largely addressed. They may still worry that any physical proximity of inspectors or equipment to sensitive information is risky, but states accepted that in the nuclear case.

\end{itemize}
Second, using chip mechanisms for verification would pose direct costs. \textbf{Certain preparations would reduce these direct costs to costs that were manageable} in nuclear arms control.

\begin{itemize}
	\item Verifying the presence and integrity of chip-based verification mechanisms would presumably require inspections. These could be implemented by adapting methods that were accepted in the nuclear case; an appendix details how this could be done with 3+ layers of defense. Back-of-the-envelope calculations suggest that, if rules’ scope were compute-intensive AI development in data centers\footnote{This would require commodity chips to not offer loopholes.}, then direct inspection costs (i.e. funding and interruptions to facilities) would be lower than or roughly similar to those which states accepted for nonproliferation M\&V.

	\item R\&D for acceptably priced hardware verification and limits on rules’ scope could theoretically reduce manufacturing and computational costs enough to keep overall costs similar to those of nonproliferation M\&V.

\end{itemize}
Third, efforts to create M\&V systems for AI could be stalled by the lack of relevant precedents and authorities. As with the above challenges, \textbf{certain preparations would reduce these cultural and legal barriers to levels that were manageable} in nuclear arms control.

\begin{itemize}
	\item Stakeholders can first create a limited M\&V system for AI, especially one with flexible authorities and scalable M\&V methods. This would lower cultural and legal barriers to a strong M\&V system\footnote{This also presumably helps ease concerns about direct costs, secrecy, and security, by letting stakeholders learn in lower-stake contexts that the downsides of an M\&V method are acceptable.}.

	\item All the strongest nuclear M\&V systems succeeded weaker systems. For example, the scope of nonproliferation M\&V expanded from just research reactors to all nuclear facilities in a state. Similarly, the U.S. and U.S.S.R. applied inspections to intermediate-range missiles before applying them to (higher-stakes) long-range missiles.

\end{itemize}
This analysis highlights that stakeholders can help enable future AI treaty verification by developing acceptable hardware verification methods and building an initial, scalable verification system. For making these preparations, Shavit  \cite{shavit2023} suggests potential next steps. The analogy of nuclear arms control suggests such steps are neither futile nor excessive; they could change verification challenges from unprecedented to historically manageable.

\newpage
\tableofcontents
\newpage

\section{Background} \label{1background}

\subsection{Introduction} \label{11introduction}

As the capabilities of AI systems grow, so do their potential benefits, as well as their safety, security, and misuse risks. Current or upcoming AI systems may greatly advance areas such as education and healthcare, but they also risk escalating military conflicts, empowering disinformation and terrorist activities, and causing large-scale accidents  \cite{brundageetal2018}\cite{bommasanietal2021}\cite{hendrycksetal2022}\cite{horowitzandscharre2021}\cite{rudnerandtoner2021}. Ambitious potential approaches to addressing these risks include \textbf{international agreements that guardrail AI development and/or deployment}  \cite{allenandwest2021}\cite{erdlyiandgoldsmith2018}, ideally complementing responsible domestic and corporate policies. For example, we could (optimistically) imagine a few countries which lead in AI reaching an enforceable agreement on the following practice: in their territories, any training run with certain high-risk characteristics\footnote{For example, one criteria could be whether the amount of compute used is above some high threshold. Such a threshold would ideally be updated in such a way that high-risk AI development is continuously covered, even if the compute requirements for high-risk AI development decrease over time due to advances in algorithmic efficiency  \cite{erdilandbesiroglu2022}.} would require the advance approval of an excellent, independent AI safety auditor. If implemented well enough, such an agreement would ensure that, across these countries\footnote{To prevent harmful AI development from states that are not parties to such an agreement, non-parties would need to be sufficiently cautious, behind, deterred, or otherwise blocked from harmful AI development.}, AI developers that cut corners on safety would not outpace sufficiently responsible ones. A similar practice could help limit the misuse potential of AI systems, and other kinds of agreements could help prevent the unintended escalation of military conflict by AI systems.

Efforts to bring about international agreements on AI would need to overcome many challenges. An especially difficult challenge could be that, for these agreements to be enforceable, good \textbf{monitoring and verification (M\&V) of compliance seems necessary}. States must be able to quickly and reliably catch defectors, or harmful defection would be much more likely.\footnote{This is because, insofar as an M\&V system is weak and under certain assumptions, state parties (i) will be less deterred from defection, (ii) will be more incentivized to defect in order to preempt others’ defection, and (iii) will be less likely to be stopped if they do defect. (One assumption here is that—due to the high levels of expert scrutiny that high-stakes M\&V systems tend to receive—a weak M\&V system will likely be perceived as such. Another assumption is that, in the absence of effective M\&V, compliance decisions will be made with the incentives of a prisoner’s dilemma or a low-trust assurance game.)} At the same time, M\&V must not be so revealing of industry or state secrets, or otherwise costly, that states would find it unacceptable.

Faced with the questions of whether effective M\&V for AI is feasible and what would help bring it about, we are fortunate to be able to learn from the experience of nuclear arms control. This is the area where most efforts to negotiate and implement ambitious M\&V systems for international security have taken place,\footnote{Nuclear arms control history involves over 50 years of international M\&V and over a dozen international agreements with verification requirements. In contrast, other areas of post-WWII nonproliferation and arms control history involve only two major international agreements with verification requirements: the Chemical Weapons Convention and the (now abandoned) Treaty on Conventional Armed Forces in Europe.} and it is a context with some similarities (though significant differences) to that of AI.\footnote{See \hyperref[thenuclearaianalogy]{an appendix} for more detailed discussion of the analogy.} Some research, e.g.  \cite{maas2019}\cite{scharreandlamberth2022}\cite{zaidianddafoe2021}, has studied nuclear arms control as a case study for AI governance, but it appears that no existing research closely investigates the M\&V aspects of this analogy. To fill that gap, this report details the history of\textbf{ M\&V in nuclear arms control} and analyzes its implications for AI treaties.

In considering lessons for the case of AI, we focus on potential rules on AI development that requires very large numbers of specialized computer chips, since verification in this context appears relatively tractable and increasingly important  \cite{bommasanietal2021}\cite{sevillaetal2022}. Concurrent and closely related work by Shavit  \cite{shavit2023} describes a potential technical approach to implementing M\&V in this context.

As a limitation, by focusing on verification, we only consider one factor in whether and how stakeholders should pursue AI treaties. Other important considerations, such as other reasons why negotiations could fail\footnote{Beyond verification problems, other reasons why negotiations could fail include partisan obstructionism, international tensions, lack of political will at high levels of governments, anti-regulation advocacy, inability to keep pace with rapid technological advances, commitment problems, issue linkage, and negotiators misunderstanding each other’s red lines.} or why a treaty may be undesirable\footnote{For example, some rules on AI activities may be undesirable due to economic or military costs.}, are beyond the scope of this report.

This report is organized as follows. Section \hyperref[1background]{1} reviews \textbf{background} context on nuclear arms control. Section \hyperref[2nuclearmvmethods]{2} describes \textbf{methods} used for nuclear arms control M\&V. Section \hyperref[3nuclearmvsystems]{3} overviews how these individual M\&V methods have been combined and implemented as \textbf{M\&V systems}. Section \hyperref[4trackrecordsofnuclearmvsystems]{4} assesses these systems’ \textbf{track records} at detecting attempted violations. Section \hyperref[5politicsofmvnegotiations]{5} describes the \textbf{politics} of the negotiations by which nuclear arms control M\&V systems were agreed on. Section \hyperref[6lessonsforaitreaties]{6} argues for \textbf{lessons }for the M\&V of international agreements on AI. Section \hyperref[7conclusion]{7} concludes. Lastly, appendices argue in more detail for various claims, and they describe a potential system for verifying the locations and integrity of AI-specialized chips.

\subsection{Three types of nuclear arms control M\&V systems} \label{12threetypesofnucleararmscontrolmvsystems}

Some context is helpful for making sense of nuclear arms control M\&V. In the early 1960s, the Cuban Missile Crisis convinced world leaders to do more to prevent nuclear apocalypse.\footnote{A retired diplomat who was involved in NPT negotiations recalls, "the 1962 crisis was the trigger that prompted a widespread change with regard to this issue. Humankind was faced with a crisis that could end in a global catastrophe. It is after that crisis the negotiations on the nuclear weapons problem began in earnest"  \cite{timerbaev2017}. An IAEA report  \cite{iaea1998} and Schelling  \cite{schelling2005} agree.}

Since then, states have implemented three clusters of nuclear arms control treaties and associated M\&V systems:

\begin{enumerate}
	\item In \textbf{horizontal nonproliferation} treaties, states that did not have nuclear weapons agreed to never make them. The most important of these treaties is the Non-Proliferation Treaty (NPT).\footnote{There are also a handful of Nuclear-Weapon-Free-Zone treaties—horizontal nonproliferation treaties that only apply to certain regions—as well as a few agreements that just sought to keep one specific state (i.e. North Korea or Iran) from developing nuclear weapons.}

	\begin{enumerate}
		\item The NPT opened for signature in 1968, when five states were recognized as having nuclear weapons. Almost all states that did not have nuclear weapons at the time joined the NPT as non-nuclear-weapon states and did not develop nuclear weapons afterward. However, four states never joined (or in one case, left) the NPT and developed nuclear weapons.
		\item The NPT requires non-nuclear-weapon states to accept an M\&V system implemented by the International Atomic Energy Agency (IAEA).
	\end{enumerate}
	\item In \textbf{U.S.-U.S.S.R./Russia nuclear arms limitation }treaties, the U.S. and U.S.S.R./Russia agreed to limit (and later, progressively reduce) the numbers or types of their deployed\footnote{In the context of nuclear arms control, "deployed" usually means "ready to be launched." For example, a nuclear weapon could be deployed if it is attached to a usable missile, but not if it is in a warehouse with no way to be launched quickly.} nuclear delivery vehicles (e.g. ICBMs) or deployed nuclear warheads (as well as, in one case, missile defense systems).\footnote{For a more detailed overview, see e.g. \href{https://www.armscontrol.org/factsheets/USRussiaNuclearAgreements}{"U.S.-Russian Nuclear Arms Control Agreements at a Glance"} from the Arms Control Association.}

	\begin{enumerate}
		\item The U.S. and U.S.S.R./Russia are widely considered to have mostly complied with these treaties.
		\item The first treaties of this type were verified largely through satellite images; later treaties also featured on-site inspections (organized by the U.S. and U.S.S.R./Russia, not the IAEA).
	\end{enumerate}
	\item In \textbf{test ban} treaties, states agreed to not carry out (certain) nuclear weapon tests.\footnote{The 1963 Partial Test Ban Treaty, a multilateral treaty, banned nuclear tests in the atmosphere, underwater, and in outer space. The Threshold Test Ban Treaty and the Peaceful Nuclear Explosions Treaty were U.S.-U.S.S.R. treaties that banned all nuclear explosions above a certain threshold of energy released. The 1996 Comprehensive Test Ban Treaty would have banned all nuclear explosions, but it requires 44 specific countries to ratify it in order for it to go into effect, and 8 of these have not ratified it.}

	\begin{enumerate}
		\item There have been no clear cases of violations of these treaties.
		\item These treaties have mostly been verified through sensors (e.g. seismic sensors) that can detect nuclear weapon tests from a long distance.

	\end{enumerate}
\end{enumerate}
Appendices elaborate on \hyperref[nwfztreaties]{regional nuclear-weapon-free zone treaties} and \hyperref[agreementswithnorthkoreaandiran]{nonproliferation agreements with North Korea and Iran}.

These M\&V systems were centered on tracking nuclear materials and nuclear delivery vehicles.

\subsection{Nuclear materials} \label{13nuclearmaterials}

The IAEA’s efforts to verify horizontal nonproliferation \textbf{focus on tracking uranium and plutonium}, certain forms of which can be made to undergo nuclear explosions. A major reason for this focus is that acquiring such weapon-usable nuclear material is the hardest step in making nuclear weapons; in contrast, nuclear bomb design and assembly are relatively simple  \cite{rosenthaletal2019}. Additionally, uranium and plutonium are relatively rare materials and emit radiation, which makes them unusually easy to track.

Inconveniently, \textbf{the production of nuclear energy involves materials and equipment that could easily be used to make nuclear weapons} (in the absence of safeguards)  \cite{ntindb}.

\begin{itemize}
	\item Uranium enrichment refers to the concentration of a specific isotope in uranium. While uranium is typically enriched above natural levels for use as nuclear fuel, enriching it to even higher levels makes it weapon-usable. Centrifuges that make fuel-usable uranium can be rearranged to make weapon-usable uranium.
	\item Plutonium is a byproduct of the reaction that takes place in nuclear reactors. After plutonium is isolated in the reprocessing of nuclear reactor products, it can be used to make nuclear weapons.

\end{itemize}
\subsection{Nuclear delivery vehicles} \label{14nucleardeliveryvehicles}

Nuclear delivery vehicles—whose numbers are restricted in various U.S.-U.S.S.R./Russia arms limitation treaties—are equipment designed to send nuclear weapons to their targets. The main such equipment is  \cite{ntinda}:

\begin{itemize}
	\item \textbf{Missiles}, which can be launched from land, from submarines, or from aircraft.
	\begin{itemize}
		\item Nuclear missiles include "ballistic missiles" (which fly like rockets) and "cruise missiles" (which fly like airplanes).
		\item Land-based missiles can be immobile (i.e. based in silos) or mobile (i.e. installed on large vehicles that can move over roads or railways).
		\item Nuclear missiles can be "tactical" (having ranges short enough to use in a battlefield) or "strategic" (having ranges long enough for the U.S. and U.S.S.R. to hit each other’s mainland, e.g. ICBMs).
		\item Missiles can have "multiple independently targetable reentry vehicles" ("MIRVs"), meaning one missile can deliver multiple warheads to multiple locations.
	\end{itemize}
	\item "\textbf{Heavy} \textbf{bombers}," which are planes that can be equipped to drop nuclear bombs.

\end{itemize}
\section{Nuclear M\&V Methods} \label{2nuclearmvmethods}

This section provides an overview of the wide range of methods used for nuclear arms control M\&V.

\subsection{Accounting and mandatory self-reporting} \label{21accountingandmandatoryselfreporting}

Verification of horizontal nuclear nonproliferation and U.S.-U.S.S.R./Russia nuclear arms limitation treaties is largely done by \textbf{verifying accounts of treaty-regulated items}  \cite{inftreaty1987}\cite{starti1991}\cite{newstart2010}\cite{rosenthaletal2019}. The idea here is that, if an agency can verify the locations of all uranium and plutonium (or, in the case of U.S.-U.S.S.R./Russia treaties, the locations of all nuclear missiles, bombers, and/or deployed warheads) in a state, then the agency can verify that no uranium or plutonium has been diverted for nuclear weapon production (or that nuclear delivery vehicles’ numbers are below treaty limits).

State parties to the NPT and to several U.S.-U.S.S.R./Russia treaties are required to regularly self-report the quantities and locations of treaty-restricted items and facilities in their territories.\footnote{For the INF Treaty, START, and New START, states are also required to report the dimensions of all restricted delivery vehicles, to facilitate their identification.} To achieve this, states require domestic facility operators to track and report on regulated items. The remainder of these treaties’ M\&V systems focus on verifying these state reports.

Self-reporting requirements can be helpful even if states submit false reports. After all, the requirements force deceptive states to tell much more detailed lies (which are easier to falsify\footnote{For example, if Russia reports how many ICBMs it has at each facility, an inspection of a single facility could reveal that the report is false. In contrast, if Russia only reported its total number of ICBMs, knowledge of many of Russia’s nuclear missile facilities would be necessary to reveal the report is false.}), and they can lead deceptive states to accidentally self-report inconsistent information (which has happened\footnote{In a report which concluded that Syria very likely attempted to build a secret nuclear reactor, the IAEA noted, "Large quantities of barite were purchased by the AECS [Syrian Atomic Energy Agency] between 2002 and 2006. Syria has stated that the material was to be used for shielded radiation therapy rooms at hospitals, without providing any supporting information. However, the end use of the barite as stated in the actual shipping documentation indicates that the material was intended for acid filtration. Additionally, the delivery of the barite was stopped at the request of the AECS after the destruction of the building at the Dair Alzour site and the remaining quantity was left undelivered"  \cite{directorgeneral2011}.}).

As another application of mandated self-reporting, some bilateral treaties require exchanges of missile flight-test data. States use other means to verify the data’s accuracy, and the data helps states verify that new types of missiles have a compliant number of reentry vehicles.

\subsection{M\&V methods at declared facilities} \label{22mvmethodsatdeclaredfacilities}

The following M\&V methods are applied only at state-declared, treaty-restricted facilities (i.e. most types of facilities in the nuclear fuel cycle for the NPT, and missile and bomber facilities for several U.S.-U.S.S.R./Russia treaties), in order to verify self-reported accounts  \cite{inftreaty1987}\cite{starti1991}\cite{newstart2010}\cite{rosenthaletal2019}.

On-site M\&V methods are implemented by \textbf{on-site inspectors}, which the IAEA implements on a regular basis for the NPT and which the U.S. and U.S.S.R./Russia have implemented with a quota system for several bilateral treaties.

Inspections tend to involve measures for mitigating concerns about espionage\footnote{A typical procedure (used in e.g. START I) is that inspected parties escort accompanying inspectors throughout their visit, control inspectors’ travel, provide some of inspectors’ equipment, and check inspectors’ equipment, as long as they do not interfere with the inspection. In turn, inspectors may check, e.g. with calibration standards, that their equipment has not been sabotaged.} and, when relevant, for mitigating concerns that banned items could be snuck out during an inspection.\footnote{In some inspections (e.g. various START I inspections), inspectors may patrol the perimeter of the inspection site, be present at exits of the site and of some structures within the site, and inspect objects and vehicles that are leaving through these exits.}

\subsubsection{On-site measurement methods} \label{221onsitemeasurementmethods}

For verifying accounts, inspectors rely heavily on \textbf{simple methods}: visual observation, counting, measurement of item dimensions (e.g. with tape measures), and tipping nuclear material cans to test their weight.

To detect more subtle violations (e.g. removal of small amounts of nuclear material from many cans, or "dummy items"), inspectors use \textbf{sensors} (weight sensors, radiation detectors, and—rarely—X-ray scanners), and IAEA inspectors also take \textbf{samples }of materials for analysis at labs. At some facilities (especially areas where human presence is unsafe), IAEA inspectors install \textbf{unattended equipment} (e.g. item counters and radiation detectors).

\subsubsection{Containment and surveillance} \label{222containmentandsurveillance}

To enable accounting verification in radioactive or otherwise inaccessible areas, and to reduce inspection frequency for stored items (in order to reduce costs), the IAEA uses containment and surveillance. That means putting nuclear materials in \textbf{tamper-indicating containers} and/or under the eye of on-site, IAEA-installed \textbf{video surveillance cameras}.

If the containers or cameras indicate that some materials were not accessed over some period of time, the IAEA can conclude that its accounting of those materials at the beginning of the period is still accurate by the end of the period.

\subsubsection{Unique identifiers} \label{223uniqueidentifiers}

Deceptive actors may try to divert or tamper with regulated equipment and then hide this by replacing the regulated equipment with a substitute. Unique identifiers, i.e. known features that are hard to (quickly) counterfeit, make it hard to create convincing substitutes.

\textbf{Serial numbers} are used as unique identifiers to help detect the diversion of missiles in a few U.S.-U.S.S.R./Russia treaties. The IAEA also uses serial numbers to track nuclear fuel rod assemblies, and it uses more sophisticated types of unique identifiers in its tamper-indicating seals to make them harder to replace with counterfeits.

\subsubsection{Design information verification} \label{224designinformationverification}

The IAEA \textbf{requires states to report information on the design} of their nuclear facilities (e.g. their floorplans). It \textbf{verifies this information through inspectors’} use of visual observation, simple length measurement tools, satellite images, and sometimes equipment that can detect modifications to a room or nearby underground rooms. These inspections are done both during and after facility construction.

Design information verification helps ensure that (i) facility designs do not make it easy to hide violations and that (ii) the IAEA can plan its inspections with accurate information on facility designs.

\subsubsection{Perimeter portal continuous monitoring} \label{225perimeterportalcontinuousmonitoring}

At a total of three (former) missile assembly facilities in each other’s territories, the U.S. and the U.S.S.R. established "perimeter portal continuous monitoring" to verify that the facility was not shipping out a restricted type of nuclear-capable missile. This meant sending 30 monitors to live and work near these facilities, continuously operating a \textbf{system for monitoring the sites’ perimeters} (without entering the facilities).

Monitors verified that objects large enough to hold banned items only exited through a few designated, monitored exits ("portals"), with advance notice. To verify that no banned items left through other exits, monitors would use a perimeter fence, a fence integrity monitoring system (or a video camera system, with motion detectors and lights), and a data processing center.\footnote{Additionally, unobstructed underground passages out of facilities were banned, obstructed ones were "subject to examination," and incoming helicopters and cranes required advance notice outside emergencies. As another precaution, monitors also had an independent power system.} Then, at portals, monitors would inspect outgoing objects with simple measurement methods\footnote{These were: visual inspection, length measurement, and/or weight sensors.} or (when necessary) an X-ray scanner, while using gates and streetlights to control vehicle flow  \cite{startiprotocoloninspectionsandcontinuousmonitoringactivities1991}.

\subsection{M\&V methods not limited to declared facilities} \label{23mvmethodsnotlimitedtodeclaredfacilities}

Complementing on-site methods, other M\&V methods are used to detect violations that occur outside declared facilities, especially the existence of secret, treaty-violating facilities or weapons  \cite{rosenthaletal2019}. Some of these methods can also uncover violations at declared facilities.

\subsubsection{National technical means} \label{231nationaltechnicalmeans}

National technical means ("NTMs") of verification are state-owned technologies used for remotely verifying compliance with treaties  \cite{gottemoeller2020}.

\begin{itemize}
	\item Nuclear arms control verifiers use \textbf{satellites} to help detect secret nuclear weapon facilities and to verify the number of various types of nuclear missiles.
	\item The U.S. and the U.S.S.R./Russia use \textbf{radar} to detect treaty-violating missile tests.
	\item The International Monitoring System is a global network of four types of sensors (three types of \textbf{acoustic wave detectors} and one type that is a \textbf{detector of certain air particles}) built to detect treaty-violating nuclear weapon tests  \cite{ctbtond}.

\end{itemize}
NTMs have been very widely used in verification, including in the first few U.S.-U.S.S.R. nuclear arms control treaties (which had no other verification measures), suggesting states are relatively open to them.\footnote{Accordingly, U.S. chief negotiator Gottemoeller  \cite{gottemoeller2020} recalls, "Noninterference with national technical verification was one of the earliest and easiest points of agreement in the New START negotiations."}

\subsubsection{Human sources} \label{232humansources}

\textbf{Whistleblowers}, \textbf{spies}, and \textbf{loose-lipped accomplices} can reveal much information about efforts to secretly violate treaties  \cite{lewis2006b}. While this is often informal, under Additional Protocols (discussed \hyperref[32mvforhorizontalnonproliferationagreements]{below}), the IAEA has wide \textbf{authority to interview} people involved in states’ nuclear programs.

\subsubsection{Intelligence sharing} \label{233intelligencesharing}

The \textbf{sharing of information} between treaty-implementing agencies and national intelligence agencies (or between multiple national intelligence agencies) can help treaty verifiers reach conclusions that would be harder to reach with more limited information. Nuclear arms control treaties have no requirements for national intelligence agencies to share their information; the process is voluntary, informal, and inconsistent.

\subsubsection{Procurement monitoring} \label{234procurementmonitoring}

Intelligence agencies \textbf{monitor nuclear-relevant trade}, since the international purchase of certain equipment components or facilities can suggest intentions to proliferate. Case studies show that methods used for this have included human sources  \cite{lewis2006b}, shipment intercepts  \cite{nti2015b}, intelligence sharing  \cite{burr2017}, and information provision requirements\footnote{In one special case, the Iran Deal, Iran was required to inform an oversight organization of a wide range of its nuclear-relevant purchases (this was called a "monitoring procurement channel").}.

\subsubsection{Other methods of unilateral information collection} \label{235othermethodsofunilateralinformationcollection}

Complementing agreed-on measures for verification, treaty parties (presumably, given their incentives and capabilities) use additional unilateral methods for verification. In addition to \hyperref[232humansources]{espionage} and \hyperref[234procurementmonitoring]{procurement monitoring}, state intelligence agencies might use \textbf{communications intercepts}  \cite{follath2015} and \textbf{open-source information}.The IAEA also uses open-source information.

\subsubsection{Challenge inspections and requests for additional information} \label{236challengeinspectionsandrequestsforadditionalinformation}

The above sources of information are often suggestive but not decisive or internationally credible. For example, satellite images might be ambiguous, and documents shared by a national intelligence agency might be fake.

To more credibly resolve ambiguities, nuclear arms control treaties often allow verifiers to request additional information (though a state might not provide it), and sometimes—most prominently, in IAEA Additional Protocols, discussed \hyperref[32mvforhorizontalnonproliferationagreements]{below}—they authorize "challenge inspections"\footnote{This is the term used for such inspections in the Chemical Weapons Convention; the IAEA uses the term "complementary access inspections."}: \textbf{inspections at nearly any location in a state’s territory}. (In contrast, other inspections only take place at certain declared, relevant sites.)

Challenge inspections can feature \textbf{environmental sampling}: taking samples (e.g. by swiping a cloth over a surface) to detect unexplained particles of nuclear material.

\subsection{Methods to boost other M\&V methods} \label{24methodstoboostothermvmethods}

In addition to methods for collecting data, M\&V often involves methods for improving how informative the collected data is  \cite{inftreaty1987}\cite{starti1991}\cite{newstart2010}\cite{rosenthaletal2019}.

\subsubsection{Equipment validation and other methods to counter specific deception tactics} \label{241equipmentvalidationandothermethodstocounterspecificdeceptiontactics}

Various methods discussed above involve inspectors using or installing certain equipment. To verify that this equipment and its data are not tampered with, inspectors typically use:

\begin{itemize}
	\item \textbf{Containment and surveillance of the equipment} itself (e.g. tamper-indicating devices on video cameras)
	\item \textbf{Data authentication}
	\item \textbf{Equipment tests} (e.g. with calibration materials)

\end{itemize}
In addition to equipment validation, M\&V systems often involve the following measures to counter specific ways a state might attempt to deceive them:

\begin{itemize}
	\item \textbf{Short-notice inspections} make it harder for states to clear out signs of non-compliance before inspectors arrive.
	\item \textbf{Random sampling} (for determining the location and timing of inspections, as well as items inspected) allows inspectors to be more efficient, while not letting states carry out violations with items they know will not be inspected.
	\item \textbf{Bans on interference with or deliberate concealment from NTMs}\footnote{These provisions ban e.g. jamming satellites and putting nets over mobile missiles  \cite{gottemoeller2020}.} allow NTMs to work more reliably.

	\item \textbf{Bans on concealment of missile flight-test data} make it harder for states to self-report fabricated flight-test data.\footnote{States could hide their flight-test data from foreign NTMs by encrypting their data, jamming foreign sensors, broadcasting their data in a narrow beam, or enclosing their data in physical capsules that fall to the ground. To deter such deception, the START I Treaty explicitly bans all of these actions, while also more generally banning "any activity that denies full access to telemetric information"  \cite{starti1991}. (As exceptions, though, this treaty allows 11 uses of flight-test data encapsulation or encryption per year.)}

\end{itemize}
\subsubsection{Methods to reduce ambiguity} \label{242methodstoreduceambiguity}

Nuclear M\&V systems sometimes involve the following requirements, which make it easier for verifiers to get relatively unambiguous information:

\begin{itemize}
	\item \textbf{Distinguishing characteristics}: certain regulated items or activities are required to be easily distinguishable.\footnote{For example, under START I, different missile types are required to have distinguishing characteristics, which helps determine missile types through inspections or satellite images. Missiles lacking these characteristics could be detected as such through visual inspection or measurement.}

	\item \textbf{Displays and exhibitions}: certain regulated items (e.g. missiles) are required to be clearly shown to inspectors or to satellites (by opening up roofs).
	\item \textbf{Location restrictions on equipment}: some equipment (e.g. missiles) is only allowed to be in certain locations.
	\item \textbf{Limits on the number (and size) of certain types of buildings or facilities}: used most in bilateral nuclear arms control, these reduce the number of inspections needed.

\end{itemize}
\section{Nuclear M\&V Systems} \label{3nuclearmvsystems}

In order to reliably detect nuclear arms control violations, the above M\&V methods must be combined and implemented. This section provides an overview of the several nuclear arms control M\&V systems that have been implemented very widely or between great powers.

\subsection{High-level thoroughness and redundancy} \label{31highlevelthoroughnessandredundancy}

At a high level, we might expect nuclear M\&V systems to work well because of their \textbf{thoroughness}. At their best, nuclear M\&V systems are designed with the aim of ensuring that \emph{any} attempted treaty violation would be quickly detected, with high enough probability\footnote{The probability of detection varies by violation. It can be hard to quantify, since non-public and non-systematic intelligence gathering methods (e.g. espionage) sometimes play large roles in M\&V. Still, in its systematic methods, the IAEA’s standard is 90\% for detecting diversions of the most sensitive nuclear materials from declared nuclear facilities.} to deter the violation  \cite{rosenthaletal2019}.

An important aspect of thoroughness—for it to be robust to design or implementation failures—is \textbf{redundancy}. Accordingly, nuclear M\&V systems use multiple layers of defense  \cite{ctbtond}.

\subsection{M\&V for horizontal nonproliferation agreements} \label{32mvforhorizontalnonproliferationagreements}

\subsubsection{IAEA safeguard systems} \label{321iaeasafeguardsystems}

The NPT explicitly requires non-nuclear-weapon state parties to accept IAEA\footnote{Founded a decade prior to the signing of the NPT, the IAEA had already been implementing nuclear safeguards on a more limited scope: verifying the peaceful use of some nuclear exports.} "safeguards" (i.e. M\&V methods, with the details to be agreed on by the state and the IAEA) on all nuclear material in the state. The IAEA mainly implements these safeguards through two systems, in accordance with two types of agreements  \cite{rosenthaletal2019}:

\begin{enumerate}
	\item \textbf{Comprehensive Safeguards Agreements ("CSAs")}: The IAEA negotiated a single template\footnote{This template is called INFCIRC/153.} which it has used as the basis for all its agreements with NPT non-nuclear-weapon states. The resulting agreements are called CSAs. CSAs are intended to (just) verify the peaceful use of nuclear materials at known nuclear facilities, rather than also detecting secret nuclear facilities.

	\item \textbf{CSAs with Additional Protocols ("APs"):} In the early 1990s, Iraq nearly made nuclear weapons by using secret nuclear facilities. In response, governments pushed for the IAEA to expand its M\&V so that it would be better at detecting such violations. The IAEA did so by negotiating a new template\footnote{This template is officially called the "Model Protocol," or INFCIRC/540.} for new agreements (called Additional Protocols), which supplement CSAs and improve the IAEA’s ability to detect secret nuclear facilities.

	\begin{enumerate}
		\item While over half of non-nuclear-weapon state parties to the NPT have now adopted APs (especially ones with significant use of nuclear materials), many have not; doing so is not generally seen as an obligation from the NPT itself.\footnote{Perhaps this is a path dependency from the historical legal interpretations of the NPT, which for a long time was only interpreted as mandating CSAs (partly since APs did not exist yet).}
	\end{enumerate}
\end{enumerate}
Besides the NPT, there have been about eight other prominent horizontal nuclear nonproliferation agreements, which apply to limited geographic regions or specific states (North Korea and Iran). Instead of having distinct M\&V systems, these agreements mainly or entirely require state parties to adopt CSAs (sometimes with APs), implemented by the IAEA.\footnote{See the \hyperref[nwfztreaties]{relevant} \hyperref[agreementswithnorthkoreaandiran]{appendices} for details and evidence.}

The IAEA reports sufficiently serious\footnote{To report potential non-compliance, the IAEA does not need to have proof that a state is diverting nuclear materials for nuclear purposes; it just needs to be unable to verify that a state is compliant. Because of that, rejection of permission to carry out a special investigation can be—and has been—a reason for reporting to the UN Security Council. In other words, the operating principle, in theory, is, "guilty until verified innocent." In practice, though, the IAEA does not report all anomalies or non-cooperation to the UN Security Council, instead appearing to reserve reporting for cases where there is relatively legible evidence of serious violations.} incidents of possible non-compliance to the UN General Assembly and the UN Security Council  \cite{rosenthaletal2019}. Then, it is up to states to decide what to do.\footnote{The NPT does not specify penalties for non-compliance  \cite{npt1968}. The template for CSAs, INFCIRC/153, does authorize the IAEA to take certain punitive measures specified in the IAEA’s statute (although these do not seem like strong deterrents against nuclear weapons development): "direct curtailment or suspension of assistance being provided by the Agency or by a member, and call for the return of materials and equipment made available to the recipient member or group of members. The Agency may also, in accordance with article XIX, suspend any non-complying member from the exercise of the privileges and rights of membership"  \cite{infcirc1531972}\cite{iaea1956}}

For its verification activities in 2022, the IAEA had a \textbf{budget of approximately \$150 million}  \cite{iaea2022e}.

\subsubsection{IAEA verification in declared nuclear facilities} \label{322iaeaverificationindeclarednuclearfacilities}

Under CSAs, the IAEA’s verification system at declared nuclear facilities works through the following process  \cite{rosenthaletal2019}:

\begin{itemize}
	\item \textbf{States report facilities:} States are required to self-report the existence and location of all facilities in their territories that hold nuclear material, except for uranium mines, uranium mills, and certain waste facilities. Once self-reported, these facilities are referred to as "declared nuclear facilities."
	\item \textbf{States report accounts:} States are required to (effectively mandate nuclear facility operators to) keep and report accounts of nuclear materials at their declared facilities. A single facility typically has multiple areas in which nuclear material stocks and flows are tracked.
	\item \textbf{The IAEA inspects facilities: }The IAEA conducts on-site inspections at declared nuclear facilities to verify the accuracy of their reported nuclear accounts.
	\begin{itemize}
		\item \textbf{The IAEA conducts inspections at the frequency it estimates to be sufficient }for identifying the diversion of nuclear materials before nuclear weapon construction can be finished. This is typically one, three, or twelve months, depending on the material.\footnote{Critics argue that this is outdated, as modern enrichment technology allows for significantly faster enrichment than when these goals were set. The IAEA does use some "Limited Frequency Unannounced Access" inspections for short-notice inspections of enrichment plants.}\footnote{Most nuclear power reactors are refueled once a year or two, and inspectors are present during this time.} In addition to regular, scheduled inspections, the IAEA also uses randomly timed inspections. Inspections come with a minimum 24-hour notice.\footnote{As is standard for international inspections, representatives of the host state are allowed to accompany inspectors (to ensure they are not spying), but they may not interfere with inspectors’ work.}

		\item To detect violations at declared facilities, inspectors use many of the methods described earlier: \textbf{\hyperref[221onsitemeasurementmethods]{on-site measurement}} (especially counting, using radiation detectors, and \textbf{taking samples} for analysis), \textbf{\hyperref[222containmentandsurveillance]{containment and surveillance}}, and \textbf{\hyperref[224designinformationverification]{design information verification}}.
		\item For \textbf{redundancy}, inspectors use combinations of measurements.\footnote{e.g., measuring both weight and contents of a container}

		\item To detect violations that involve diverting large amounts of nuclear material from a few containers, inspectors make quick, rough measurements of many containers; to detect violations that involve diverting \emph{small} amounts of nuclear material from \emph{many} containers, inspectors apply more sensitive methods to a sufficiently large random sample of containers.\footnote{If a state diverted some material from many nuclear containers, random selection allows the IAEA to have a high chance of identifying diversions while only having to precisely measure a small fraction of all containers. There is a tradeoff between the IAEA’s desired probability of detecting diversion and the number of measurements needed, with returns to further measurements diminishing quickly. The IAEA has explicitly decided to make this tradeoff such that its detection probability is 90-95\% for the most sensitive nuclear material.}

		\item To verify flows, the IAEA requires nuclear facility operators to declare when they have received certain materials\footnote{these are "mailbox declarations"} and to hold them for a specified time. The IAEA verifies this with short-notice randomized inspections.

		\item Traditionally, the IAEA developed safeguard agreements based just on the characteristics of each facility. Over the last decade, it has increasingly adopted "state-level approaches": deciding safeguard implementation with more consideration of the broader context in a state.\footnote{For example, if the IAEA believes a state does not have plutonium reprocessing plants (which are needed to develop plutonium-based nuclear bombs), that may be a reason for the IAEA to put fewer resources (e.g., less frequent inspections) on safeguarding of plutonium in the state’s nuclear waste, while putting more resources on safeguarding uranium at enrichment plants. As of the end of 2017, the IAEA had developed 62 state-level safeguards approaches, with plans for more.}

		\item In addition to on-site inspections, the IAEA also uses unattended and remote safeguards: video cameras and machines that do automated counting and measurement of nuclear materials.
	\end{itemize}
	\item \textbf{The IAEA analyzes inspection data and resolves anomalies: }When safeguards show inconsistencies or odd findings that (in aggregate) are significant\footnote{The IAEA sets a goal for safeguards to identify all diversions of "significant quantities" of nuclear materials, defined based on its estimates of how much is necessary to produce a single nuclear weapon. These estimates include: 8 kg for plutonium, 25 kg for U-235 in highly enriched uranium, 75 kg for U-235 in low enriched uranium, and 10 t of natural uranium. Critics argue these estimates are overly high, and nuclear weapons could be made with nuclear material that is smaller by a factor of e.g., 2-8.}, the IAEA by default first seeks clarification from the relevant state.

	\begin{itemize}
		\item It can do this by, e.g., requesting additional information, redoing verifications (sometimes by shutting down a process line while inventories are re-counted), or requesting to carry out "special Investigations" (which include access beyond that which the IAEA would normally have).

	\end{itemize}
\end{itemize}
When APs supplement CSAs, they do not greatly change the above process at declared facilities—see \hyperref[howadditionalprotocolschangemvprocessesatdeclarednuclearfacilities]{the relevant appendix} for details.

\subsubsection{IAEA verification of the absence of undeclared nuclear facilities} \label{323iaeaverificationoftheabsenceofundeclarednuclearfacilities}

We can break down the process of detecting undeclared facilities into two steps:\footnote{Some M\&V mechanisms are well-suited for (1), while others are well-suited for (2). More precisely, some mechanisms are cheap to use at scale yet have high false positive rates, while other mechanisms are costly yet more reliable. An efficient strategy can be to use the former mechanisms to identify suspected locations and the latter mechanisms to determine whether these locations hold undeclared nuclear facilities. An analogy from epidemiology is the use of "pool testing" to cheaply test many individuals for a disease, followed by more precise (yet costly) testing of a smaller number of individuals who are highlighted by the pool testing.}

\begin{enumerate}
	\item Finding evidence suggesting that a state might have undeclared nuclear facilities (potentially at a specific location), and
	\item Resolving suspicions about suspected undeclared nuclear facilities.\footnote{As discussed earlier, suspicions do not need to be positively resolved for the IAEA to report non-compliance to the UN Security Council. Still, reducing the frequency of false alarms seems likely useful for ensuring that states take alarms seriously.}
\end{enumerate}
\textbf{Mechanisms for identifying suspect locations or states:}

\begin{itemize}
	\item \textbf{Unofficially, the IAEA identifies suspect locations mainly through voluntary tips from national intelligence agencies} (and perhaps also from whistleblowers). These tips from intelligence agencies appear to be irreplaceable for the IAEA’s ability to identify potential undeclared nuclear facilities.\footnote{As discussed in \hyperref[theroleofintelligenceagenciesinidentifyingundeclarednuclearfacilities]{the relevant appendix}, these claims are supported by strong trends in expert opinion and case studies, as well as the existence of a plausible mechanism that might cause them to be true.}\footnote{Intelligence agencies’ tips are also available in the context of states that only have CSAs, but they play a vital role in verification only in the context of states that also have APs. This is because it is only for these latter states that the IAEA has the authority to act on tips through investigation; IAEA accusations of non-compliance based entirely on state tips would not be credible.}\footnote{We may wonder: When national intelligence agencies already know about some undeclared nuclear facility, what is the point of the IAEA investigating it? Perhaps IAEA investigations enable findings of non-compliance with the NPT to have international credibility (and thus bring about international penalties), which intelligence agencies (with their opaque methods and national bias) may be unable to do on their own.}

	\begin{itemize}
		\item Intelligence agencies have not published much information on their methods for detecting undeclared nuclear facilities, but some reports suggest their \hyperref[235othermethodsofunilateralinformationcollection]{methods} include: spying, monitoring nuclear trade, analyzing satellite images, and using other open-source information  \cite{lewis2006b}\cite{harelandbenn2018}.
		\item The IAEA also conducts its own analyses to identify suspect locations, but these are limited by the IAEA’s limited sources of information; compared to the high bar of the IAEA’s processes at declared facilities, the IAEA has no similarly comprehensive, independent process for reliably finding undeclared facilities.\footnote{As reasons to conclude this, the IAEA does not publicly describe any such process, it is very unclear how IAEA safeguards agreements (which are public) could provide the authorities needed for such a process, and the IAEA has a \hyperref[41nptmvsystemstrackrecords]{spotty track record in identifying undeclared facilities}.}

	\end{itemize}
	\item Expanding IAEA abilities under CSAs, states that have also adopted APs are required to provide the IAEA with information on additional nuclear buildings and activities  \cite{rosenthaletal2019}.\footnote{These are:

	\begin{itemize}
		\item Non-nuclear buildings on the same "site" as nuclear facilities
		\item Nuclear material mines, mills, and certain waste facilities
		\begin{itemize}
			\item Information about these partly helps indirectly, by helping identify diversions of nuclear material.
		\end{itemize}
		\item Nuclear R\&D activities that do not use nuclear materials
		\item Relevant equipment manufacturing activities that do not use nuclear materials (e.g. centrifuge manufacturing)
		\item Decommissioned nuclear facilities	\end{itemize}} Presumably, whether or not states actually self-report on these facilities, these reporting requirements help the IAEA detect secret nuclear facilities.\footnote{If a state self-reports on one of the above facilities, then the IAEA can investigate activities there and discover that it supplies (or is supplied by) an undeclared nuclear facility. For example, if the IAEA found that some nuclear waste had less plutonium than expected, this may indicate that some of the plutonium had been secretly transported to an undeclared reprocessing facility. (The IAEA could also find that the declared facility is itself doing unreported processing of nuclear materials, although this would not be a case of finding an undeclared nuclear facility.)
    If a state does not self-report on one of the above facilities but the IAEA somehow discovers it (e.g., through intelligence agency tips followed by an inspection with \hyperref[236challengeinspectionsandrequestsforadditionalinformation]{environmental sampling}), this would strongly indicate that the state has (other) undeclared nuclear facilities or activities, since states generally do not have peaceful reasons to have undeclared nuclear facilities. (In contrast, discovering that a state did not declare a facility would be less suspicious if the state had not been required to declare the facility.)}
\end{itemize}
\textbf{Mechanisms for resolving suspicions about suspect locations or states} (and their limitations)  \cite{rosenthaletal2019}:

\begin{itemize}
	\item \textbf{CSAs grant the IAEA very limited means for confirming or disconfirming its suspicions about undeclared nuclear facilities.}\footnote{See \hyperref[theiaeasinabilitytoverifytheabsenceofundeclarednuclearfacilitiesundercomprehensivesafeguardsagreements]{the relevant appendix} for justification of this claim.}

	\begin{itemize}
		\item Recognizing this, the IAEA does not report the absence of undeclared nuclear facilities in states that only have CSAs.
	\end{itemize}
	\item When they supplement CSAs,\textbf{ APs boost the IAEA’s ability to identify and investigate suspected undeclared nuclear facilities}. APs authorize the IAEA to \hyperref[236challengeinspectionsandrequestsforadditionalinformation]{investigate suspect locations} in certain ways:
	\begin{itemize}
		\item For all the locations that APs require states to self-report (listed in a footnote under the last sub-heading), APs also grant the IAEA "complementary access" to these locations. This means that, with certain qualifications (especially for fully private activities), APs grant the IAEA \textbf{non-routine inspection access to a very wide range of nuclear-relevant locations}, with 24 hours’ notice.\footnote{At inspections, inspectors may use any of the safeguards usually available to them (described in other sections), although accounting is infeasible because states are not required to keep nuclear materials accounts at these additional locations.}

		\item APs also grant the IAEA \textbf{complementary access to any location in a state}, with a couple of qualifications.\footnote{These qualifications are:

	\begin{itemize}
		\item Access is only given so that the IAEA can carry out "environmental samples and, in the event the results do not resolve the question or inconsistency at the location specified by the Agency [...], utilization at that location of visual observation, radiation detection and measurement devices, and, as agreed by [the state] and the Agency, other objective measures"  \cite{infcirc5401997}.
		\item The state is permitted to be unable to provide such access, in which case the state "shall make every reasonable effort to satisfy [IAEA] requirements, without delay, at adjacent locations or through other means…"  \cite{infcirc5401997}.	\end{itemize}}

	\end{itemize}
\end{itemize}
\subsection{M\&V for U.S.-U.S.S.R./Russia nuclear arms limitation agreements} \label{33mvforusussrrussianucleararmslimitationagreements}

\subsubsection{M\&V for SALT I Agreements and SORT} \label{331mvforsaltiagreementsandsort}

Several U.S.-U.S.S.R./Russia nuclear arms limitation agreements had little or no officially authorized M\&V:

\begin{itemize}
	\item \textbf{SALT I Agreements:} The earliest U.S.-U.S.S.R. nuclear arms control agreements (the SALT I interim agreement and associated Anti-Ballistic Missile Treaty, which capped offensive nuclear missiles and missile defense systems, respectively) \textbf{only had \hyperref[231nationaltechnicalmeans]{national technical means}} ("NTMs") as formally authorized M\&V methods  \cite{salti1972}\cite{abmtreaty1972}.
	\begin{itemize}
		\item The Soviets rejected proposals for inspections, due to concerns over espionage.
	\end{itemize}
	\item \textbf{SORT:} The 2002 Strategic Offensive Reductions Treaty ("SORT"), aka the Moscow Treaty, had \textbf{no formal verification methods of its own}. However, SORT mostly overlapped with START I, which had an extensive M\&V system, so in practice there were some M\&V mechanisms allowing the U.S. and Russia to get information about whether the other was moving toward compliance with SORT.
	\begin{itemize}
		\item The U.S. intelligence community concluded that, outside of the period where SORT would overlap with START I, it could not confidently verify Russian compliance, but Russia would likely comply regardless due to the costs of maintaining its nuclear arsenal. Also, SORT provisions only officially applied for one day. After outlining this context, arms control expert Jeffrey Lewis writes that \textbf{SORT "isn’t arms control but domestic political theater"}  \cite{lewis2004}.

	\end{itemize}
\end{itemize}
\subsubsection{M\&V for the INF Treaty, START I, and New START} \label{332mvfortheinftreatystartiandnewstart}

In the INF Treaty, START I, and New START, which have collectively constituted the majority of U.S.-U.S.S.R./Russia nuclear arms reduction treaties since 1987, verification tends to work in the following way  \cite{inftreaty1987}\cite{starti1991}\cite{newstart2010}:

\begin{itemize}
	\item \textbf{Each party \hyperref[21accountingandmandatoryselfreporting]{shares and updates data} on where all its nuclear delivery vehicles are} and when any of them are destroyed, while also sharing missile flight-test ("telemetry") data.
	\item \textbf{Each party verifies the other’s shared data through various \hyperref[221onsitemeasurementmethods]{on-site inspections} coupled with \hyperref[231nationaltechnicalmeans]{NTMs}} and (presumably) \hyperref[235othermethodsofunilateralinformationcollection]{other unilateral intelligence gathering methods} (e.g. \textbf{espionage}).
	\begin{itemize}
		\item Inspections mainly use simple methods, e.g. counting and length measurement.\footnote{These treaties use annual quotas for inspections (unlike the IAEA), ranging from 10 to 28.}

		\item Inspectors only inspect declared facilities; the task of discovering secret facilities is left to other methods (e.g. satellites).
	\end{itemize}
	\item Additionally, two of these treaties use \hyperref[225perimeterportalcontinuousmonitoring]{perimeter portal continuous monitoring} at a few missile assembly facilities, and the treaties involve further \hyperref[24methodstoboostothermvmethods]{measures that make it easier} for inspections and NTMs to verify compliance: restrictions on equipment locations, distinguishing characteristics, limits on the number of (size-limited) on-site buildings, and \hyperref[223uniqueidentifiers]{unique identifiers}.

\end{itemize}
Differences between these treaties’ verification systems are discussed in a footnote\footnote{Perhaps the most significant differences are:

	\begin{itemize}
		\item The INF Treaty was a total ban on intermediate-range (500-5500 km) missiles, while START I and New START required reductions in strategic offensive arms (i.e. long-range nuclear weapons).
		\item Specifically in the context of the INF Treaty, maintaining or innovating on the banned missiles would require flight tests at banned ranges, so flight-test monitoring could be enough to detect treaty violations.
		\item The INF Treaty and START I M\&V included \hyperref[225perimeterportalcontinuousmonitoring]{perimeter portal continuous monitoring} at a few facilities, while the New START did not, reportedly because this monitoring was expensive and not very useful  \cite{gottemoeller2020}.
		\item To streamline the M\&V process, New START consolidated the many types of inspections of START I into just two (more extensive) types of inspections. It also replaced complex counting rules (i.e. rules about how many warheads particular types of missiles would be counted as having) with simpler rules, which worked through inspectors directly counting the number of warheads on missiles at inspections.	\end{itemize}}, and more details on implementation can be found in the \hyperref[2nuclearmvmethods]{earlier section} on M\&V methods.

\subsection{M\&V for nuclear weapon test bans} \label{34mvfornuclearweapontestbans}

\textbf{All 4 nuclear weapon test ban treaties are verified mainly through sensors meant to detect distant nuclear explosions}, and \textbf{2 of the treaties also have provisions for on-site inspections.}

\subsubsection{National technical means and the IMS} \label{341nationaltechnicalmeansandtheims}

For over two decades, the Comprehensive Test Ban Treaty Organization ("CTBTO") has been building the International Monitoring System—\textbf{a global network of ~300 sensors designed to detect nuclear explosions}  \cite{ctbtond}. This system is operational despite the relevant treaty—the Comprehensive Test Ban Treaty ("CTBT")—not having entered into force.\footnote{The CTBT called for the Comprehensive Test Ban Treaty Organization (which the CTBT established) to begin construction of the International Monitoring System so that it would be ready by the time the treaty entered into force. The CTBT currently looks far from entering into force; by treaty provision, a large number of specific countries must all sign it for it to enter into force, and various have not. Still, the U.S. and some other states have been providing funding to support the construction of the monitoring system, and (legally) the treaty does not need to be in force for states to be able to host sensors on their own territories.} The system consists of four types of sensors, which detect acoustic waves or air particles that are indicative of nuclear explosions.\footnote{The four types of stations are: seismic, hydroacoustic, infrasound, and radionuclide sensors.} In addition to the sensor stations, the CTBTO operates a system to transmit sensor data to a data processing center, which shares raw and analyzed data with member states. Analysts can use sensors’ data to infer the location of nuclear tests.\footnote{To help enforce nuclear weapon test bans, analysts do not just need to know when a test happened; they also need to know where it happened. The location of a nuclear test can be determined from acoustic wave sensors similarly to how one can use seismic sensors to locate the epicenter of an earthquake. And the location of a nuclear test can be roughly determined from air particles by having a model of global wind patterns and running the model backwards in time.}

In addition to sensors overseen by the CTBTO, states and some private organizations have their own relevant sensors  \cite{ntindc}. These were the main basis of verification for nuclear test ban treaties that preceded the CTBT.

The CTBTO had a \textbf{budget of approximately \$100 million} allocated for verification in 2020  \cite{ctbto2021}\footnote{As of December 2022, this is the most recent annual report published by the CTBTO. Previous years’ reports show that the CTBTO’s 2020 budget was similar to that of preceding years, despite the COVID-19 pandemic. \$100 million is an approximation of the reported \$131,320,100 multiplied by 81\%, the percentage reportedly allocated to verification.}.

\subsubsection{On-site inspections} \label{342onsiteinspections}

The Peaceful Nuclear Explosions Treaty—a U.S.-U.S.S.R. treaty that was ratified in 1990 and caps the yield (i.e. energy) of peaceful nuclear explosions\footnote{Peaceful nuclear explosions hypothetically might be used, for example, for mining or building canals.}—authorizes \textbf{on-site inspections at the sites of peaceful nuclear explosions} that are sufficiently close to the permitted threshold, and it also establishes \textbf{data exchange requirements} for parties to inform each other about these tests. Depending on the details of the explosions, inspectors may be able to use \textbf{visual observation, photo cameras, a local seismic network, electrical equipment for yield determination, and a geologist field kit}  \cite{pnetreaty1976}.

The CTBT, a ban on all nuclear explosions, would authorize \textbf{\hyperref[236challengeinspectionsandrequestsforadditionalinformation]{challenge inspections}} if it entered into force. These would be conducted by the CTBTO\footnote{Inspections would require the approval of at least 30/51 members of the CTBTO’s Executive Council. These members are state parties.}  \cite{ctbt1996}. Inspectors would be authorized to carry out \textbf{visual observation}, use various types of \textbf{sensors}, and take \textbf{samples} (including by drilling)\footnote{Potential methods to gain evidence about whether a nuclear explosion took place would include: detecting radioactive materials, detecting cavities and rubble zones, and detecting seismic aftershocks.}  \cite{ctbtprotocol1996}.

\section{Track Records of Nuclear M\&V Systems} \label{4trackrecordsofnuclearmvsystems}

Having reviewed how nuclear M\&V systems were implemented, we turn to their track records, focusing on the frequency of false negatives in detecting violations. As we will see, \textbf{within each of the three main categories of nuclear arms control treaties, the strongest widely implemented M\&V system has had zero known major failures} and (up to) a few known attempts at serious violations, all of which the system detected. \textbf{However, weaker predecessors of these systems had some known failures}.

\subsection{NPT M\&V systems' Track Records} \label{41nptmvsystemstrackrecords}

First, we review the strengths of the track record of \hyperref[32mvforhorizontalnonproliferationagreements]{IAEA verification}, which include that \textbf{Comprehensive Safeguards Agreements ("CSAs") boosted by Additional Protocols ("APs") have had no known, major failures.}

\begin{itemize}
	\item As far as is publicly known, IAEA safeguards have never completely failed; \textbf{no state has ever acquired nuclear weapons while being party to the Non-Proliferation Treaty} (which involves signing a CSA with the IAEA) (Bleek, 2017; UN ODA).\footnote{The events that have been closest to NPT non-nuclear-weapon states getting nuclear weapons have been:

	\begin{itemize}
		\item Nuclear-weapon states have stored nuclear weapons in non-nuclear-weapon state parties to the NPT.
		\item North Korea publicly (although procedurally questionably) withdrew from the NPT several years before acquiring nuclear weapons.
		\item While formally under the NPT, Iran has been moving toward having the capacity to quickly produce nuclear weapons, e.g., by enriching uranium to nearly weapons-grade levels.	\end{itemize}}
	\begin{itemize}
		\item For context, a Belfer Center paper counts 7 non-nuclear-weapon states as having pursued nuclear weapons while being parties to the NPT  \cite{bleek2017}\footnote{These have been: Iran, Iraq, Libya, North Korea, Syria, South Korea, and Yugoslavia.}.

	\end{itemize}
	\item Although dozens of states now have nuclear facilities and CSAs began to be implemented about 50 years ago, \textbf{no state has attempted to divert a significant quantity of nuclear material from a facility that was under CSA safeguards}.\footnote{See \hyperref[detailsontrackrecordsofnucleararmscontrolmv]{an appendix} for evidence.}

	\item APs, which began to be implemented about 25 years ago to address CSAs’ weakness at detecting secret nuclear facilities, are nearly untested in their ability to detect secret nuclear weapons development activities; \textbf{there are no known cases of states attempting to build or operate secret nuclear fuel-cycle facilities while under an AP}. AP safeguards successfully detected Iran’s engagement in some undeclared nuclear activities, the details of which remain unclear.
	\begin{itemize}
		\item There is almost no known case of a state pursuing a nuclear weapons program while under a CSA with an AP.\footnote{This conclusion is based on the dataset of Bleek  \cite{bleek2017}, which lists when all known state pursuits of nuclear weapons took place, considered together with an IAEA "Status list"  \cite{iaea2022d}, which lists when states signed APs. However, that is a much shorter—and thus much less informative—track record than the track record of CSAs; CSAs began to be implemented in the early 70s, while APs began to supplement them in the late 90s.}

		\item The one known exception is Iran\footnote{Iran claims to not be pursuing a nuclear weapon, but this claim is hard to reconcile with Iran’s high uranium enrichment levels, its self-reported technical ability to make the bomb, and Iran’s multiple (no-longer-)secret nuclear facilities and activities.}\footnote{APs also helped the IAEA identify past undeclared nuclear activities in South Korea and Egypt, but these activities had been small-scale research activities and not parts of state pursuits of nuclear weapons  \cite{gao2005}.}, which openly expanded its nuclear activities and (as determined by the IAEA) also engaged in undeclared activities involving nuclear material, while implementing a CSA with an AP\footnote{This happened in the several years when the Trump Administration had withdrawn the U.S. from the Iran Deal but Iran had not yet ended its implementation of its AP. (Iran’s expanded nuclear activities have continued after Iran stopped implementing its AP, but these post-AP activities are outside the scope of analyzing the technical track record of CSAs with APs.)}. Although Iran’s limited compliance has kept the IAEA from uncovering the details of these activities, the IAEA was able to discover that they occurred\footnote{Using its CSA and AP authorities, the IAEA requested and received "complementary access" for inspectors to conduct \hyperref[236challengeinspectionsandrequestsforadditionalinformation]{environmental sampling} at three suspected nuclear sites in Iran (at least some of which the IAEA received tips about from states). Analyzing these samples, the IAEA found that all three sites held anthropogenic uranium particles, which Iran struggled to explain  \cite{directorgeneral2022}. It is not yet clear what the delay was between these undeclared activities taking place and the IAEA discovering them.}  \cite{directorgeneral2022}.

		\item Even when it is not dealing with unique cases like Iran, the IAEA often faces and accepts significant delays in resolving uncertainties about the existence of undeclared nuclear facilities. For example, regarding the year 2021\footnote{One might wonder if this example is an outlier due to the COVID-19 pandemic, but this appears to not be the case; the numbers were similar (62/131) the year before the pandemic began  \cite{iaea2020c}.}, for 60 of the 132 states with CSAs and APs in force, the IAEA reported that "[e]valuations regarding the absence of undeclared nuclear material and activities for each of these States remained ongoing"\footnote{Because these investigations were ongoing, the IAEA did not conclude that these states did not have undeclared nuclear facilities.}  \cite{iaea2022c}.
	\end{itemize}
\end{itemize}
However, the IAEA’s track record is far from perfect; \textbf{CSA safeguards not complemented by APs repeatedly missed states’ construction of secret nuclear facilities} (which they were not designed to detect).

\begin{itemize}
	\item Iran, Iraq, Libya, and Syria all pursued nuclear weapons by building secret nuclear facilities while under CSAs, but CSA safeguards were not enough for the IAEA to notice  \cite{nti2015a}\cite{nti2015b}\cite{nti2018b}\cite{nti2020}. Similarly, the IAEA failed to notice Yugoslavian R\&D and South Korean international purchases that pursued nuclear weapons while these states were under CSAs  \cite{potteretal2000}\cite{burr2017}.
	\item As exceptions to the above trend, CSA safeguards have occasionally identified banned activities outside of what they are primarily designed to identify.
	\begin{itemize}
		\item In North Korea, following the state’s adoption of a CSA, initial IAEA inspections found that North Korea had previously processed more plutonium than it claimed. This finding kicked off (ultimately unsuccessful) diplomatic efforts to keep North Korea from getting the bomb  \cite{nti2018a}.
		\item In Syria, the IAEA concluded that a destroyed facility had "very likely" been an undeclared nuclear reactor (though this was several years after Israel bombed it)  \cite{boardofgovernors2011}.\footnote{Unlike with North Korea, this finding did not kick off extensive diplomacy, as Syria had recently become occupied with a civil war.}

	\end{itemize}
\end{itemize}
\subsection{Bilateral Nuclear Arms Control M\&V systems' Track Records} \label{42bilateralnucleararmscontrolmvsystemstrackrecords}

For all U.S.-U.S.S.R./Russia nuclear arms limitation agreements that were mutually ratified and contained M\&V measures, \textbf{neither party is widely considered to have attempted the most serious violation possible}: maintaining numbers of strategic arms far above their permitted levels. In this sense, \hyperref[33mvforusussrrussianucleararmslimitationagreements]{these M\&V systems} are untested, though the lack of known violation attempts is some evidence for these systems’ (perceived) reliability.

\textbf{Across the three bilateral arms control agreements with \hyperref[332mvfortheinftreatystartiandnewstart]{extensive M\&V systems}} (i.e. the INF Treaty, START I, and New START), there has been \textbf{only one state allegation of an attempt at secret, serious noncompliance}: 26 years after the INF Treaty entered into force, the U.S. accused Russia of discreetly developing and flight-testing a missile that violated the treaty. (The U.S. consequently withdrew from the treaty.)

\begin{itemize}
	\item The timing of the U.S. allegations is ambiguous in its implications about the reliability of U.S. M\&V, but a tentative conclusion is that U.S. M\&V (especially NTMs) was adequate.
	\item Otherwise, the U.S. has mostly confirmed the U.S.S.R./Russia’s compliance, and Russia has made several relatively minor or tenuous allegations of noncompliance.

\end{itemize}
In bilateral arms control treaties that had \hyperref[331mvforsaltiagreementsandsort]{more limited M\&V systems} (e.g. just NTMs), the U.S. and the U.S.S.R. made several accusations of non-compliance (though not of massive non-compliance except for an unratified treaty). Given the ambiguity of these cases and the limited role of M\&V, they tell us less about relevant M\&V systems’ effectiveness.

See \hyperref[detailsontrackrecordsofnucleararmscontrolmv]{an appendix} for more details and evidence on the above claims about parties’ allegations.

\subsection{Nuclear Test Ban M\&V Systems' Track Records} \label{43nucleartestbanmvsystemstrackrecords}

\textbf{There has been no clear, prominent case of a nuclear test ban being violated.} Even the Comprehensive Test Ban Treaty, which has not entered into force, is considered to have been followed by signing parties.\footnote{See \hyperref[detailsontrackrecordsofnucleararmscontrolmv]{an appendix} for more details.}

The most significant potential exception is the "Vela incident"; \textbf{there is mixed evidence about whether a Partial Test Ban Treaty state tested a nuclear weapon in 1979}  \cite{cohenandburr2020}. Even if this violation did happen, though, it would have been before the construction of the \hyperref[341nationaltechnicalmeansandtheims]{International Monitoring System}, so such a violation would face a stronger monitoring system today.

The International Monitoring System successfully detected all six of North Korea’s known nuclear tests (confirmed by the state’s announcements), providing some evidence for the system’s effectiveness  \cite{ntindc}\cite{ctbto2022}.

\subsection{Limitations} \label{44limitations}

The public would not know about any nuclear weapons development activities that were sufficiently well-kept secrets. However, \textbf{there are some reasons to expect that there have not been not many such well-kept nuclear secrets}:

\begin{itemize}
	\item Some of the factors that most plausibly incentivize states to have nuclear weapons programs—deterrence and prestige—also incentivize states to inform other states and the public when they succeed at developing, testing, or stockpiling nuclear weapons. This suggests that most nuclear weapons programs that have succeeded are widely known.\footnote{Of the 10 states that are widely considered to have acquired nuclear weapons, 2—Israel and South Africa—were not quick to publicly announce them  \cite{nti2014}\cite{nti2015c}\cite{ourworldindata2022}.}

	\item Changes in leadership or in state incentives can motivate states to reveal formerly secret programs.\footnote{This happened to varying degrees in, for example, Yugoslavia, Brazil, South Africa, Iraq, and Libya  \cite{potteretal2000}\cite{burr2017}\cite{nti2006}\cite{nti2015a}\cite{nti2015b}\cite{nti2015c}.}
\end{itemize}
There is also significant uncertainty due to incentives for states to spread misinformation about their own or others’ secret nuclear weapons activities.

\section{Politics of M\&V Negotiations} \label{5politicsofmvnegotiations}

This section describes aspects of the politics of negotiations in which M\&V systems were agreed on, focusing on the negotiations of the templates for Comprehensive Safeguards Agreements ("CSAs") and Additional Protocols ("APs"), the main templates for \hyperref[32mvforhorizontalnonproliferationagreements]{IAEA M\&V agreements}. Due to limitations in available sources\footnote{e.g., this is not based on detailed public sources on informal negotiations on M\&V implementation details, due to the apparent absence of such sources.}, this overview is relatively light and non-comprehensive.

\subsection{Incentives in M\&V negotiations} \label{51incentivesinmvnegotiations}

Consistently with negotiators’ incentives, the records and outcomes of M\&V negotiations suggest that these negotiations involved significant pressures to do the following:

\begin{itemize}
	\item \textbf{Ensure effectiveness}\footnote{The U.S. Chief negotiator of New START writes, "First, the verification regime of any arms control treaty must be effective"  \cite{gottemoeller2020}. Similarly, a declassified 3-volume history of IAEA M\&V negotiations concludes that a "fundamental" negotiating objective of the U.S. in INFCIRC/153 (i.e. CSA template) negotiations, which was "[t]o a considerable degree [...] shared by other participants," was to "preserve the integrity and effectiveness of the IAEA safeguards system"  \cite{internationalenergyassociateslimited1984}.}

	\begin{itemize}
		\item This is reflected in M\&V systems’ \hyperref[31highlevelthoroughnessandredundancy]{thoroughness and redundancy}.
	\end{itemize}
	\item \textbf{Protect state and commercial secrets}\footnote{An IAEA historian writes, "States feared—and continue to fear—that inspectors might conduct industrial espionage [...]"  \cite{roehrlich2018}. Accordingly, an IAEA report explains, "The IAEA has put in place elaborate arrangements to ensure that safeguards information remains confidential and to prevent unauthorized disclosures. [...] Despite the very large quantity of safeguards information handled by the IAEA since 1970, there has not been any substantiated case of such a disclosure nor any complaint on this score by the government of any State in which safeguards are applied"  \cite{iaea1998}.}

	\begin{itemize}
		\item Inspections include a wide range of measures for addressing this concern.\footnote{Some example precautions are: inspectors’ equipment being inspected when it reaches the inspected country, inspectors being escorted throughout their visit, perimeter monitors not being allowed to enter the facility they monitor, video cameras not transmitting live feeds remotely, and missile reentry vehicles being covered with pliable covers during inspection.}

		\item States that are technical leaders in some industries appeared to be especially concerned about IP risks.\footnote{For example, Japan advocated for especially strict non-disclosure policies for IAEA inspectors  \cite{internationalenergyassociateslimited1984}.}

	\end{itemize}
	\item \textbf{Limit disruptions and financial costs}
	\begin{itemize}
		\item Inspections have been limited and streamlined to address this concern.\footnote{The U.S. chief negotiator of New START writes, "We had heard loud and clear from our military services that they were concerned about the costs of elimination procedures and the operational interruptions that were happening because of inspections"  \cite{gottemoeller2021}. She also writes that these concerns informed negotiations, motivating a streamlining of inspections  \cite{gottemoeller2020}. Similarly, an IAEA negotiating history states that, "Cost was a big concern to G-77 states in particular. They consistently worried that the addition of new measures would come at the expense of technical assistance to developing countries"  \cite{rosenthaletal2010a}.}

		\item States appeared to be more concerned over M\&V being applied to the specific supply chain steps they led in.\footnote{A history of IAEA negotiations explains  that major uranium and thorium producers opposed safeguards on uranium mining and ore processing  \cite{internationalenergyassociateslimited1984}.}

	\end{itemize}
	\item \textbf{Limit other security threats}
	\begin{itemize}
		\item START I allows roughly its entire, extensive M\&V system to be arbitrarily suspended for "operational dispersals" of nuclear forces. Analysts explain, "In view of the central importance of preserving the survivability of their strategic forces, the Parties were unwilling to place any restrictions on the number, frequency, or duration of operational dispersals. [...] [H]owever, the Parties specify that such operational dispersals shall only be conducted for national security purposes in time of crisis when a Party considers it necessary to act to ensure the survivability of its strategic forces [...] [and only] rarely"  \cite{federationofamericanscientists1998}.
	\end{itemize}
	\item \textbf{Preserve national industrial competitiveness}
	\begin{itemize}
		\item In initial NPT negotiations as well as later AP negotiations, nuclear-weapon states volunteered to accept IAEA safeguards on their own commercial nuclear plants. This move was reportedly critical for easing "widespread concerns" that IAEA safeguards "would place non-nuclear-weapon States at a commercial and industrial disadvantage in developing nuclear energy"  \cite{rosenthaletal2010a}\cite{rosenthaletal2010b}\cite{rosenthaletal2019}.
	\end{itemize}
	\item \textbf{Keep compliance feasible}
	\begin{itemize}
		\item This reportedly motivated minimum-notice requirements for inspections.\footnote{According to an IAEA negotiating history, "A practical reason behind many of these proposals [minimum notice requirements, e.g. minimum 2-hour notice before inspections] was the difficulty of making arrangements for access to buildings on short or no-notice that do not in fact contain nuclear material and to which the state and facility representatives accompanying the Agency inspectors might not themselves have access." Additionally, India pointed to the difficulty of collecting information from industry  \cite{rosenthaletal2010b}.}

	\end{itemize}
	\item \textbf{Observe privacy rights}
	\begin{itemize}
		\item IAEA inspectors lack unconditional access to arbitrary locations.\footnote{An IAEA negotiating history explains that, "states were apprehensive about the intrusiveness of the access and about their ability to provide the access in light of their constitutions. This was especially the case under the earlier Secretariat drafts for access for environmental sampling, which could be essentially anywhere in the state for any purpose"; the scope of complementary access was qualified to address this concern  \cite{rosenthaletal2010b}.}

	\end{itemize}
	\item \textbf{Avoid passing new legislation}\footnote{One expert reviewer disagrees with the assessment that this was a significant obstacle.}

	\begin{itemize}
		\item States repeatedly objected to IAEA M\&V proposals on the grounds that they would require changes to national laws and regulations  \cite{rosenthaletal2010a}.
	\end{itemize}
	\item \textbf{Appease idiosyncratic stakeholders}
	\begin{itemize}
		\item The U.S. chief negotiator of New START suggests the treaty includes telemetry data sharing because senators mistakenly considered it useful for verification.\footnote{The U.S. chief negotiator of the New START, writes that, although the treaty updated its predecessor’s (i.e. START’s) verification system such that telemetry measures became obsolete ("we did not need telemetry measures to confirm compliance"), some senators were "particularly determined that telemetry measures should be part of the new treaty because they had been such a central player in the success of START"  \cite{gottemoeller2020}\cite{gottemoeller2021}. The treaty ended up including telemetry measures.}

	\end{itemize}
	\item \textbf{Limit partiality among states}
	\begin{itemize}
		\item Nuclear verification tends to be consistent across states, and safeguards agreements emphasize the use of "objective methods"  \cite{infcirc1531972}.\footnote{A nuclear nonproliferation expert writes that, "In the initial development of safeguards, a requirement to avoid discrimination among participating states was met by adopting uniformity in safeguards application"  \cite{carlsonetal1999}.}

	\end{itemize}
	\item \textbf{Respect national sovereignty}
	\begin{itemize}
		\item As discussed \hyperref[64managingculturalandlegalbarriers]{below}, concerns over national sovereignty appear to have motivated various early limitations on IAEA safeguards.\footnote{An IAEA historian writes, "States [...] often perceive the [nuclear safeguards] system as a threat to national sovereignty"  \cite{roehrlich2018}. In this vein, an IAEA report states that, early in the development of the IAEA’s verification system, "The concepts of short notice and unannounced inspections, now increasingly important features of IAEA safeguards, would have been regarded as inadmissible infractions of national sovereignty"  \cite{iaea1998}.}
	\end{itemize}
\end{itemize}
In the end, agreed-on IAEA safeguards were a mountain of \textbf{compromises, tending to achieve high reliability while involving various limitations to reduce costs} for host states.\footnote{As a few examples, the point in the supply chain when safeguards would start, the frequency and intensity of inspections, the amount of advance notice required for inspections, and the limits on complementary access inspections were all settled as compromises between parties who wanted more and less strict measures. As another example, the following provision in CSAs expresses a typical compromise position: "The Agency shall require only the minimum amount of information and data consistent with carrying out its responsibilities under the Agreement"  \cite{infcirc1531972}.}

\subsection{Why Comprehensive Safeguards Agreements were not designed to detect secret nuclear facilities} \label{52whycomprehensivesafeguardsagreementswerenotdesignedtodetectsecretnuclearfacilities}

Perhaps the biggest failure in nuclear M\&V has been one discussed \hyperref[41nptmvsystemstrackrecords]{above}: that CSAs were not designed to detect secret nuclear facilities. Seemingly emboldened and enabled by this, a handful of NPT non-nuclear-weapon states ran secret nuclear weapons programs while under CSAs, and usually they (especially the most advanced programs) relied on secret nuclear facilities.

How did CSA negotiations come to leave such a massive gap in their M\&V system? Experts propose the following explanations (though typically with little to no citation/evidence) for why CSAs had very limited capacity for detecting undeclared nuclear facilities: negotiators had tended to think that:

\begin{itemize}
	\item Secret nuclear facilities would be detected and voluntarily reported on by national intelligence agencies  \cite{iaea1998}\cite{carlsonetal1999}\cite{carlsonetal2020}\cite{rosenthaletal2010a};
	\item Establishing a self-contained nuclear fuel cycle would be too technically difficult for most states  \cite{carlsonetal1999}\cite{timerbaev2017};
	\item Inspectors having far-reaching access to investigate potential violations was politically unacceptable  \cite{iaea1998}\cite{carlsonetal2020}; and
	\item There were no good available methods for the IAEA to detect undeclared facilities  \cite{carlsonetal2020}.

\end{itemize}
Considering this alongside the fact that later fixes (Additional Protocols) required ratification from each state party, \textbf{it appears that the CSA M\&V system has been highly flawed because negotiators made fragile assumptions, built insufficient flexibility into CSAs, and were insufficiently proactive in responding to changes in the risk landscape}.\footnote{Some assumptions were fragile in the sense of being false (the assumption that national intelligence agencies would reliably detect and report on secret facilities, which did not happen with Iraq), while some were fragile in the sense of being true at the time but not for long (the other assumptions discussed above). The slowness and non-universality of AP adoption suggests the system was built with insufficient flexibility, though one may reasonably worry (as negotiators at the time did) that flexibility could have been used to weaken the system  \cite{internationalenergyassociateslimited1984}. And presumably one could have known before the near-miss with Iraq (discussed later) that proliferation had become easier and the Overton Window on inspections had shifted (from states having gained experience with inspections), though admittedly no secret nuclear programs of NPT non-nuclear-weapon states had been known to get far before Iraq’s.}

\subsection{The impact of salient failure} \label{53theimpactofsalientfailure}

Despite its initial weaknesses (discussed above), \textbf{the IAEA’s safeguards system was substantially strengthened in response to a salient failure}. This failure was Iraq’s nearly successful secret nuclear weapons program, which was discovered only through the U.S.-led coalition’s victory against Iraq in the First Gulf War  \cite{grossietal2015}\cite{rosenthaletal2019}.\footnote{Incidents with North Korea and South Africa at around the same time are also sometimes credited with contributing to this change  \cite{rosenthaletal2019}. In North Korea, the IAEA (with tips from states) found that North Korea had submitted false reports and was likely hiding secret facilities. In South Africa, the government’s initiative to cooperatively dismantle its own nuclear program helped the IAEA learn about its capacity and the requirements for verifying the absence of undeclared nuclear facilities. However, discussions of Additional Protocols tend to give the incident with Iraq much more emphasis (often not even mentioning the events with North Korea and South Africa), suggesting the discovery of Iraq’s nuclear program was a much stronger motivator, as one might expect (since it was the clearest case of IAEA failure). For instance, several years before he became Director General of the IAEA, Grossi stated, "what Chernobyl was to safety, Iraq was to safeguards, and 9-11 was to security"  \cite{grossietal2015}.}

After finding in 1991 that Iraq nearly made nukes under IAEA inspectors’ noses (the secret program operated in buildings adjacent to declared facilities), states successfully pushed for the IAEA to expand its role to not just safeguarding declared nuclear facilities but also detecting secret nuclear facilities. The IAEA determined that its existing agreements gave it authorities it had not been using—such as requiring earlier provision of design information and doing \hyperref[236challengeinspectionsandrequestsforadditionalinformation]{environmental sampling}—and it began using these authorities. Additionally, the IAEA developed and agreed with dozens of states on APs, which brought the IAEA greatly \hyperref[323iaeaverificationoftheabsenceofundeclarednuclearfacilities]{improved authorities} for verifying the absence of undeclared nuclear facilities.

The \hyperref[41nptmvsystemstrackrecords]{above} discussion of the IAEA’s systems and track records suggests APs have largely been successful.

\section{Lessons for AI Treaties} \label{6lessonsforaitreaties}

\subsection{Qualified optimism} \label{61qualifiedoptimism}

Having reviewed the implementation, track records, and politics of monitoring and verification in nuclear arms control, we now turn to consider their implications for AI treaties.

This section primarily argues for the following conclusion: \textbf{with certain preparations, the foreseeable challenges of one potential form of AI treaty verification }(specifically, hardware-based verification of treaties setting rules on highly compute-intensive AI development)\textbf{ would mostly be challenges that were \hyperref[4trackrecordsofnuclearmvsystems]{successfully} addressed in nuclear arms control.} The main \textbf{preparations needed} to prevent worse challenges are:

\begin{enumerate}
	\item \textbf{Developing privacy-preserving, secure, and acceptably priced methods for verifying the compliance of hardware}, given inspection access; and
	\item \textbf{Establishing an initial, incomplete verification system that is relatively easy to improve} when opportunities arise, because it has flexible authorities and scalable precedents.

\end{enumerate}
These are tall orders, but their potential suggests qualified optimism and plausible directions to move toward. More concretely, Shavit  \cite{shavit2023} describes some near-term policies and technical research questions that could serve as steps toward (1) and (2).

The following sections expand on the following argument for the above conclusion:

\begin{itemize}
	\item First, \textbf{consider this potential, high-level approach to verifying AI treaties}: require and verify that computer chips used for highly compute-intensive AI development have built-in mechanisms that enable verification, then use these mechanisms to verify compliance.
	\begin{itemize}
		\item Ongoing research suggests this may be technically feasible, even in privacy-preserving and efficient ways. Shavit  \cite{shavit2023} provides a technical description of one way chips could be used to verify compliance.
		\item For concreteness, one example of a chip mechanism that would help enable verification is a tamper-evident log of chip activity.
		\item Chip-based approaches to verification cannot address all important risks from AI. Still, chip-based verification may be unusually promising specifically in the context of highly compute-intensive AI development, as other drivers of AI advances—algorithms and data—are harder to track.
	\end{itemize}
	\item M\&V politics in the nuclear case suggest the above approach to verification \textbf{will mostly encounter challenges from: secrecy and security concerns; direct costs; and cultural and legal difficulties.}\footnote{One may object that this approach—considering political problems that arose in the nuclear case—neglects technical problems that may arise in the case of AI. However, these technical problems, as discussed by e.g. Shavit  \cite{shavit2023}, tend to be about enabling privacy-preserving and efficient verification (rather than about enabling any sort of verification), and those problems are captured in the substantial preparations that this report suggests are important.}

	\begin{itemize}
		\item This list covers all of the sources of objections to M\&V proposals identified in the \hyperref[5politicsofmvnegotiations]{earlier section} on politics, except for pressures to: ensure effectiveness, preserve national industrial competitiveness, keep compliance feasible, observe privacy rights, appease idiosyncratic stakeholders, and limit partiality among states.These potential concerns are de-emphasized below because they appear less likely to be major concerns or are already addressed by the preparations discussed next.\footnote{To elaborate:

	\begin{itemize}
		\item The importance of maintaining effectiveness and feasibility as other costs are reduced is implicitly assumed in the below discussion.
		\item A coalition participating in an AI treaty could maintain its industrial competitiveness through e.g. multilateral export controls and technical collaboration.
		\item Privacy rights could be observed by limiting the scope of inspections to data centers (rather than, say, households) and using secrecy-preserving M\&V methods.
		\item While specific challenges in appeasing idiosyncratic stakeholders and maintaining impartiality are hard to foresee, there are no immediately apparent reasons why these challenges would be worse in the AI case than in the nuclear case.	\end{itemize}}
	\end{itemize}
	\item \textbf{Substantial preparations would reduce each of these challenges to a difficulty that was manageable in the nuclear case} (that is, some nuclear arms control M\&V system faced a similar or greater difficulty, yet the system was adopted and had a strong track record).
	\begin{itemize}
		\item The next three sections argue for this claim in more detail.

	\end{itemize}
\end{itemize}
\subsection{Managing secrecy and security concerns} \label{62managingsecrecyandsecurityconcerns}

AI chip users may oppose verification due to concerns that (1) disclosure of AI chips’ locations, (2) inspections of AI chips, and (3) certain design features on AI chips would expose sensitive data and software to spies and saboteurs. Similar concerns could arise in the context of chip-making machines. \textbf{Certain preparations would reduce these secrecy and security concerns to concerns that were manageable} in nuclear arms control.

\begin{itemize}
	\item The importance of (1) for verification poses the challenge of needing states to disclose sensitive facilities’ locations. This \hyperref[21accountingandmandatoryselfreporting]{was manageable} in nuclear arms control.
	\begin{itemize}
		\item Nearly all states agreed to disclose the locations of their nuclear energy facilities for IAEA verification.
		\item The U.S. and U.S.S.R. agreed to share the locations of their nuclear weapon bases with each other for INF Treaty, START, and New START verification.
	\end{itemize}
	\item If stakeholders \textbf{develop privacy-preserving and secure}\footnote{That is, "secure" in the broad sense of not creating severe vulnerabilities to national security.}\textbf{ methods for using (2) and (3) to verify AI chips’ compliance}, then AI chip users’ concerns here would be largely addressed.

	\begin{itemize}
		\item States may still worry that physical proximity of inspectors or equipment to sensitive information is risky (even when no specific risks are apparent), but this concern \hyperref[22mvmethodsatdeclaredfacilities]{was manageable} in the nuclear case.
		\begin{itemize}
			\item For example, with a wide range of precautions, states accepted video surveillance and inspectors near sensitive centrifuge designs, and the U.S. and U.S.S.R./Russia allowed each other’s inspectors to be physically near their sensitive missile and warhead design information.
			\item As an example of a precaution that allows inspectors proximity but not access to sensitive information, in New START inspections, the front ends of missiles "would be opened up but covered with a soft or pliable cover so that objects [reentry vehicles] could be counted without revealing their technical characteristics"  \cite{gottemoeller2020}.

		\end{itemize}
	\end{itemize}
\end{itemize}
\subsection{Managing direct implementation costs} \label{63managingdirectimplementationcosts}

Hardware-based verification would involve costs from (1) implementing, (2) verifying the presence of, and (3) using chip mechanisms that enable verification. \textbf{Certain preparations would reduce these direct costs to costs that were manageable} in nuclear arms control.

\begin{itemize}
	\item (2) would presumably require inspections, which could be implemented by adapting \hyperref[2nuclearmvmethods]{methods} used for verifying accounts of nuclear items.
	\begin{itemize}
		\item This adaptation of nuclear materials accounting to AI chip accounting would be feasible; \hyperref[howverificationofaichipaccountscouldbeimplemented]{an appendix} details how it could be done with 3+ layers of defense against potential violations.
		\item Back-of-the-envelope calculations (in \hyperref[backoftheenvelopecalculationsofdirectinspectioncosts]{an appendix}) suggest that, if rules’ scope were highly compute-intensive AI development in data centers (meaning commodity chips would need to not offer loopholes), then direct costs of inspections (both the inspections’ funding and the disrupted economic activity) would be lower than or very roughly similar to those which states accepted for nonproliferation M\&V.
	\end{itemize}
	\item There would also be manufacturing and computational costs.
	\begin{itemize}
		\item To address these, \textbf{R\&D for acceptably priced hardware verification} and limits on rules’ scope could theoretically reduce these costs by enough to make overall costs lower than that which states accepted for nonproliferation M\&V.
	\end{itemize}
	\item AI development increasingly relies on large numbers of highly specialized chips, making it plausible that a treaty with a narrow scope would still mitigate some important risks.

\end{itemize}
\subsection{Managing cultural and legal barriers} \label{64managingculturalandlegalbarriers}

Efforts to create M\&V systems for AI could be stalled by the lack of relevant precedents and legal authorities. As with the above challenges, \textbf{certain preparations would reduce these cultural and legal barriers to levels that were manageable} in nuclear arms control.

\begin{itemize}
	\item Stakeholders can \textbf{first create a limited M\&V system for AI}, especially with flexible authorities and scalable M\&V methods. This would lower cultural and legal barriers to a strong M\&V system, which can then be created when opportunities to improve the limited system arise.\footnote{This also presumably helps ease concerns about direct costs, secrecy, and security, by letting stakeholders learn in lower-stake contexts that the downsides of an M\&V method are acceptable.}

	\item \textbf{All the strongest M\&V systems in the nuclear case were created in the above way, incrementally}, and some historians consider this critical, at least for the creation of the IAEA’s current strongest system.
	\begin{itemize}
		\item A history of IAEA safeguards  \cite{iaea1998} writes that, "there was much initial resistance to the application of IAEA safeguards. Thus the first, incomplete but complex, safeguards system covered only [...] the research and experimental reactors of the day." Over the 60s, the scope of IAEA safeguards evolved to cover all declared nuclear facilities in a state. Then, in the 90s, the scope of IAEA safeguards \hyperref[53theimpactofsalientfailure]{again extended} in many states (after the failure with Iraq) to also cover undeclared nuclear facilities.
		\begin{itemize}
			\item Another history highlights a concern that motivated incremental development: that more ambitious safeguards "would not be generally acceptable to states until further experience with IAEA safeguards was gained"  \cite{internationalenergyassociateslimited1984}.
		\end{itemize}
		\item IAEA M\&V system improvements in the 90s came most quickly and widely when there was more precedent and authority to make them  \cite{iaea1998}\cite{rosenthaletal2010a}.
		\begin{itemize}
			\item Early in IAEA safeguards’ development, according to an IAEA history, "The concepts of short notice and unannounced inspections, now increasingly important features of IAEA safeguards, would have been regarded as inadmissible infractions of national sovereignty"  \cite{iaea1998}.
			\item Some changes (e.g. the use of new technologies at inspections) could be adopted by majority vote, while other changes (e.g. expanded scope of inspections) required ratification by state legislatures; the latter changes have taken years longer and remain less widely applied.
		\end{itemize}
		\item U.S.-U.S.S.R./Russia treaties adopted increasingly intrusive M\&V for increasingly high-stakes aims.
		\begin{itemize}
			\item While the first treaties only had \hyperref[231nationaltechnicalmeans]{national technical means} as authorized M\&V methods, the INF Treaty expanded to \hyperref[332mvfortheinftreatystartiandnewstart]{inspections} for intermediate-range missiles, and the START I brought inspections to strategic (i.e. long-range) missiles.
		\end{itemize}
		\item States strengthened nuclear test ban M\&V by building \hyperref[341nationaltechnicalmeansandtheims]{many more remote sensor stations}. Meanwhile, efforts that faced higher legal barriers stalled; the Comprehensive Test Ban Treaty has still not been ratified by the number of states it requires to enter into force.

	\end{itemize}
\end{itemize}
\section{Conclusion} \label{7conclusion}

Over the last half-century, states have developed a wide range of methods and thorough systems to verify compliance with nuclear arms control agreements. The strongest widely applied verification system for each of three types of nuclear arms control treaties has had no known major failures, though weaker predecessors had significant failures. Verification systems were built to maintain effectiveness while addressing many other concerns.

This history suggests qualified optimism for the prospects of verifying agreements on AI, at least for guardrailing highly compute-intensive AI development; with certain preparations, major foreseeable problems would be reduced to ones that were manageable in the nuclear case.

\section*{Acknowledgements} \label{acknowledgements}

This research was primarily done as an independent contractor with OpenAI’s Policy Research Team. Views expressed do not necessarily represent the views of OpenAI.

I am especially grateful to Jade Leung for her support throughout this research, from guidance on research question selection to feedback on drafts. I am also thankful to the broader compute governance field, particularly Yonadav Shavit, for informing my thinking on hardware-based verification; to John Carlson, Warren Stern, and Kuhan Jeyapragasan for their generous feedback; and to OpenAI for their support.

\bibliographystyle{unsrt}  
\bibliography{references}  \newpage
\appendix

\section{The nuclear-AI analogy} \label{thenuclearaianalogy}

M\&V systems for AI would face some similar challenges as (some) M\&V systems for nuclear arms control, including:\footnote{See Zaidi and Dafoe  \cite{zaidianddafoe2021} for broader discussion of the strengths and limitations of this analogy.}

\begin{itemize}
	\item \emph{Dual-use equipment and facilities:} Much of the equipment and facilities that could be used to violate an agreement can also be used for legitimate purposes\footnote{Nuclear facilities and uranium can be used to peacefully make nuclear power, and nuclear missile facilities can store treaty-compliant numbers and types of missiles. AI-specialized hardware and data centers can be used to train and deploy treaty-compliant AI systems or other applications.}, so M\&V must be able to catch late-stage misuse of relevant equipment and facilities.

	\item \emph{Sensitive information:} Dual-use equipment and facilities involve sensitive information\footnote{Equipment relevant to nuclear arms control involves R\&D of centrifuges, missiles, and bombers, as well as the numbers and locations of nuclear arms. Equipment relevant to AI treaties might include private data, as well as AI and semiconductor R\&D.} even if they are being used in compliance with the treaty.

	\item \emph{Regulation of both government and corporate activity:} Governments and businesses perform activities that (may) need restrictions for effective M\&V.\footnote{The main case of corporate regulation in the nuclear case is that nonproliferation M\&V applies to many private nuclear energy companies.}

	\item \emph{Bilateral or multilateral negotiations:} Treaty scope is not necessarily limited to either just bilateral or just multilateral treaties.\footnote{In the case of AI, it is unclear which kinds of treaties (if any) will be feasible or desirable. Fortunately, nuclear arms control offers precedents for bilateral as well as multilateral treaties.}

	\item \emph{Need for high robustness:} States might come to see defection as highly valuable, so a reliable M\&V system might need to be robust against major state efforts to hide violations.
	\item \emph{Accounting:} Verified accounting (of uranium in one case, and of high-end, AI-specialized chips in the other case) can help with treaty verification.

\end{itemize}
However, there are also major differences between these verification challenges, including:

\begin{itemize}
	\item \emph{Degree of perceived risk:} Nuclear arms control M\&V was negotiated after clear demonstrations of the risks posed by nuclear weapons, while some potential AI risks are currently more speculative.
	\item \emph{Efficacy of environmental sampling:} The use of centrifuges to produce weapons-grade uranium scatters unique particles that can be detected from some distance; there are no obvious analogues for AI.
	\item \emph{Verification of information technology use:} M\&V for AI may need to be able to catch certain defections just based on (limited) access to source code, AI hardware, and/or ML models. Nuclear arms control M\&V has not had to do that; it offers no obvious analogues to software or hardware-centered verification.
	\item \emph{Supply chain concentration:} The supply chain of high-end computer chips is highly concentrated  \cite{khan2021}, while uranium sources, their processing equipment, and nuclear facilities are relatively decentralized. Still, in both cases, there are challenging steps in the supply chain.

\end{itemize}
Additionally, it is not clear whether certain other factors are similarities or differences, including:

\begin{itemize}
	\item \emph{Scale of verification activities needed:} The amount and scope of infrastructure and equipment that need to be inspected for nuclear arms control is low enough for verification to be considered affordable; it is unclear if the same will be true for AI.
	\item \emph{Amount of sensitive information needed to verify compliance: }Nuclear arms control agreements are verified without inspectors getting access to much of the valuable R\&D information involved (i.e. R\&D of centrifuges, missiles, and bombers); it is unclear whether similarly IP-protecting M\&V will be feasible for AI.

\end{itemize}
\section{M\&V in Nuclear-Weapon-Free-Zone treaties and in agreements with North Korea and Iran} \label{mvinnuclearweaponfreezonetreatiesandinagreementswithnorthkoreaandiran}

This appendix argues that Nuclear-Weapon-Free-Zone ("NWFZ") treaties and nonproliferation agreements with North Korea and Iran are mainly or entirely verified by \href{sometimes with APs)](\#3-2-m\&v-for-horizontal-nonproliferation-agreements-17}{CSAs (sometimes with APs)} implemented by the IAEA, rather than by distinct M\&V systems.

\subsection{NWFZ Treaties} \label{nwfztreaties}

Five treaties ban a wide range of nuclear weapon activities, including developing nuclear weapons, among state parties in a compact geographic region. Their primary verification mechanisms are standard IAEA safeguards.

\begin{itemize}
	\item Three treaties explicitly specify that state parties must adopt CSAs with the IAEA (or equivalent agreements), and one explicitly specifies that they must also adopt APs.\footnote{These are the treaties covering the South Pacific, Africa, and Central Asia. The last of these is the one that requires APs.}

	\item The treaties covering Latin America and Southeast Asia are less precise in their verification requirements,\footnote{The Latin America Nuclear Weapons Free Zone Treaty states: "Each Contracting Party shall negotiate multilateral or bilateral agreements with the International Atomic Energy Agency for the application of its safeguards to its nuclear activities"  \cite{treatyoftlatelolco1967}. The Southeast Asia Nuclear Weapon Free Zone Treaty states: "Each State Party which has not done so shall conclude an agreement with the IAEA for the application of full scope safeguards to its peaceful nuclear activities…"  \cite{bangkoktreaty1995}.} but they seem to be interpreted as also mandating (at least) CSAs with the IAEA as their primary verification systems.\footnote{Evidence for the claim that these agreements are interpreted as mandating a CSA (ever since CSAs were developed):

	\begin{itemize}
		\item The IAEA’s country-specific fact sheets  \cite{iaea2022b}, which list safeguards agreements for each country, show that various involved countries (e.g. Mexico, Peru, Guatemala, Brazil, Thailand, Cambodia, Singapore, Vietnam, Laos, Philippines, Indonesia, and Malaysia) have all lacked distinct NWFZ-specific agreements in addition to their NPT agreements (as we might expect if the agreements were different) ever since the IAEA developed CSAs. (Some countries in Latin America, such as Mexico, had NWFZ-specific agreements before CSAs were developed.)
		\item In some relevant cases (e.g., Brazil), states and the IAEA have explicitly clarified that CSAs fulfill a country’s NWFZ obligations  \cite{quadripartiteagreement2000}.
		\item Discussion of the IAEA having distinct safeguards for implementing NWFZ treaties is not prominent in descriptions of IAEA safeguards, including the sources cited in this report.	\end{itemize}}

\end{itemize}
Supplementing these IAEA safeguards, each of these treaties other than the Central Asian one also establishes a new regional, international organization. These organizations are tasked with helping implement the agreements, partly by carrying out a few verification mechanisms\footnote{These are: exchanges of reports (although nothing is specified to be nearly as extensive as reports that the IAEA requires) and requests for clarification when there are odd findings. The South Pacific and Southeast Asia treaties also authorize these regional international organizations to carry out special on-site inspections at any suspected location  \cite{treatyofrarotonga1985}\cite{bangkoktreaty1995}.}. None of these mechanisms are beyond the authority of IAEA safeguards, so these organizations at most add redundancy and regional legitimacy to standard IAEA verification.

\subsection{Agreements with North Korea and Iran} \label{agreementswithnorthkoreaandiran}

Nations have reached a few agreements in response to worries that particular nations were pursuing nuclear weapons.\footnote{These agreements tend to involve the rogue state agreeing to accept IAEA inspections and to end certain nuclear activities (often including ones that are usually legal but facilitate later nuclear weaponization, e.g. certain uranium enrichment, plutonium reprocessing, and R\&D activities) in exchange for economic or security incentives (e.g. lifted sanctions, oil, and assurances that nuclear weapons would not be used against them).} The most prominent of these (near-)agreements appear to have been\footnote{All three of these agreements were abandoned, often openly. Still, the latter two were complied with for years, so they likely slowed down North Korea and Iran’s nuclear weapons programs, respectively.}: the Joint Declaration of South and North Korea on the Denuclearization of the Korean Peninsula, the U.S.-D.P.R.K. Agreed Framework, and the Joint Comprehensive Plan of Action (JCPOA, i.e. the "Iran Deal").

These agreements tend to not break new ground in terms of M\&V mechanisms. The Joint Declaration would have involved bilateral inspections, but states failed to reach agreement on its implementation. The Agreed Framework just involved agreement to \hyperref[32mvforhorizontalnonproliferationagreements]{CSAs with the IAEA}. And, while the JCPOA involved several unusual verification measures, these were mostly \hyperref[322iaeaverificationindeclarednuclearfacilities]{standard monitoring mechanisms} applied more frequently or at more types of facilities. Specifically, under the JCPOA  \cite{rosenthaletal2019}:

\begin{itemize}
	\item There was one novel monitoring mechanism: Iran would have a monitored procurement channel, meaning it would have to notify (and get approval from) the UN Security Council (acting mostly through a commission) to engage in a wide range of nuclear-related, international economic activities (e.g. equipment imports, training);
	\item Iran would indefinitely adopt an Additional Protocol\footnote{This is the only one of these provisions that would apply indefinitely. The JCPOA’s other agreed-on verification mechanisms were set to expire in 10-25 years.};

	\item The IAEA would verify that some nuclear reactor would be reconstructed in a way that made its products less suitable for nuclear weapon production; and
	\item The IAEA would extend its usual safeguards to cover additional types of facilities.\footnote{Specifically, the IAEA would continuously monitor centrifuge production areas, excess centrifuge storage areas, uranium mines, and uranium mills, and it would monitor heavy water storage and production areas, through measures including (for some of these facilities) containment \& surveillance methods, item counting, item numbering, and/or daily inspections.}

\end{itemize}
\section{How Additional Protocols change M\&V processes at declared nuclear facilities} \label{howadditionalprotocolschangemvprocessesatdeclarednuclearfacilities}

Compared to non-nuclear-weapon states that have only adopted CSAs, states that have also adopted APs have fairly similar types of safeguards in their declared nuclear facilities. After all, APs are largely designed to improve the IAEA’s capacity to identify \emph{undeclared} nuclear facilities.

Still, there are several notable differences in safeguards at known facilities  \cite{rosenthaletal2019}; in contrast to states that have only adopted CSAs, states that have also adopted APs…

\begin{itemize}
	\item …receive relaxed safeguards ("integrated safeguards") at declared facilities, if and after the IAEA has confidently concluded that the state has no undeclared nuclear facilities.\footnote{The motivation is that, if a state has no nuclear facilities to which it can divert nuclear material, there is less risk that it will quickly develop nuclear weapons from diverted nuclear material. In integrated safeguards, safeguards are relaxed in that timeliness and detection probability goals are lowered, which leads to inspections being done less frequently, as well as to fewer samples being analyzed. These changes may have been largely implemented to avoid having APs raise safeguard implementation costs.}

	\item …agree to grant IAEA inspectors access with 2 hours notice (or less, "in exceptional circumstances") for carrying out inspections that were authorized under CSAs. The previous requirement was 24 hours.
	\item …agree to enable the IAEA to non-routinely verify that nuclear material is not being diverted from uranium mines, mills, or certain wastes (not covered by CSAs).\footnote{More precisely, states that also have APs agree to inform the IAEA of the locations and activities of these operations and to grant "complementary access": access for inspectors, upon request with 24 hours’ notice. In this access, inspectors may use most of the safeguards discussed \hyperref[32mvforhorizontalnonproliferationagreements]{above}, although they have neither the information nor the authority to verify nuclear materials accounts. (States are not required to keep detailed accounts of these operations, and the Model Protocol states that the IAEA  "shall not mechanistically or systematically" try to verify information about these additional operations  \cite{infcirc5401997}.)}\footnote{See the next appendix on why special inspection rights under CSAs were inadequate.}

	\item …agree to inform the IAEA about quantities of nuclear materials that were previously small enough to be exempted (separately from Small Quantities Protocols).

\end{itemize}
\section{The IAEA's inability to verify the absence of undeclared nuclear facilities under Comprehensive Safeguards Agreements} \label{theiaeasinabilitytoverifytheabsenceofundeclarednuclearfacilitiesundercomprehensivesafeguardsagreements}

Overall, CSAs give the IAEA very limited abilities to verify the absence of undeclared nuclear facilities  \cite{rosenthaletal2019}.

Under CSAs, the IAEA has access to the following sources of information\footnote{Additionally, the IAEA could conclude that a state has undeclared nuclear facilities by detecting that some nuclear material had been diverted from peaceful purposes and the material was not accounted for in any declared facility.}\footnote{The IAEA only began using satellite imagery in the 1990s, following failures to detect undeclared nuclear facilities, although it considers them to have been authorized by the original CSAs  \cite{iaea2021b}.}\footnote{It may be convenient for the IAEA to be able to draw conclusions from the premise that all uranium originating from a state’s nuclear mines or mills is accounted for. However, under CSAs, that is not feasible, as CSAs explicitly do not establish safeguards (e.g., accounting) at uranium mines or mills, and perhaps international uranium flows would be difficult to track.}, which can help it identify locations with a higher-than-baseline chance of hosting secret nuclear facilities:

\begin{itemize}
	\item Voluntary reports from third parties, especially national intelligence agencies
	\item States’ self-reporting on their own nuclear facilities
	\item Satellite images (e.g., images of a building whose size, location, heat\footnote{Heat can be identified in satellite images if the satellites sense infrared radiation (which all hot objects emit), rather than just sensing visible light. Alternatively, heat can be identified in satellite images by looking for rooftops where snow cover does not build up in the winter.}, security, and installations are consistent with being a uranium enrichment plant  \cite{panda2018})

	\item Other open-source information (e.g., local news reports)

\end{itemize}
Typically, all of the above sources of evidence are just suggestive\footnote{Voluntary reports from third parties could be dismissed as misinformation or disinformation  \cite{iaea2022a}, state self-reporting of nuclear facilities could leave out undeclared facilities, and satellite images or other open-source information could have non-incriminating explanations.}; if they rouse the IAEA’s suspicions about some location, the IAEA still needs more definitive evidence to back up confident accusations of non-compliance. However, CSAs only authorize the IAEA to use the following methods for resolving suspicions about undeclared facilities:

\begin{itemize}
	\item \emph{Special inspections:} Officially, under CSAs, the IAEA may make special inspections (potentially at locations that host undeclared facilities) if it considers other sources of evidence insufficient for verifying compliance. However, in practice, the IAEA has a restrictively high bar for conducting special inspections: special inspections are widely considered appropriate only when the IAEA already has credible evidence of a safeguards violation\footnote{As evidence for this claim, a report by the Institute for Science and International Security explains, "The IAEA does not often call for a special inspection—this is reserved for extreme situations where a particularly egregious safeguards violation is suspected and where the member state has demonstrated a lack of cooperation"  \cite{albrightandbrannan2010}. A paper by the director of the nuclear non-proliferation program of a major think tank at least roughly agrees: "Seoul, Tokyo, and Washington consider IAEA monitoring—even enhanced by special inspections—to be inadequate because of the limitations of the IAEA’s special inspection authority. Special IAEA inspections can be undertaken, for example, only after credible evidence of a safeguard violation has been presented, a requirement that will make such inspections highly unusual and create a political "threshold" to their use"  \cite{spector1992}. (The paper was published in 1992, when APs had not been implemented, so it is presumably about CSAs without APs.) A history of negotiations also agrees  \cite{rosenthaletal2010a}.}. In line with this, the IAEA rarely conducts special inspections\footnote{For example, when Israel bombed a building in Syria and news outlets speculated that it had held a nuclear reactor, the IAEA did not carry out a special inspection  \cite{albrightandbrannan2010}.}.

	\item \emph{Requests for additional information:} For example, the IAEA may ask that a state provide explanations or documentation about some construction activities. These requests have sometimes been useful despite the potential for deception; states have sometimes responded in ways the IAEA was able to falsify\footnote{For example, the IAEA  \cite{directorgeneral2011} found inconsistencies in additional information that Syria provided on some suspected materials, reporting that, "Syria has stated that [some] material was to be used for shielded radiation therapy rooms at hospitals [...]. However, the end use of the [material] as stated in the actual shipping documentation indicates that the material was intended for acid filtration." As another example, Iran informed the IAEA that there had been no activity at some suspected location over a particular time period; the IAEA  \cite{directorgeneral2022} reported that this was "inconsistent with the [IAEA’s] observations through the analysis of commercially available satellite imagery."}, and states have sometimes refused to respond at all\footnote{For example, the IAEA  \cite{directorgeneral2011} reports that Syria did not respond to some of their requests for information and documentation about a suspected facility. As another example, the IAEA  \cite{directorgeneral2022} reports that, when it asked Iran about suspected locations, "Iran provided no answers."}, suggesting that it can be difficult for states to craft credible cover stories.
\end{itemize}
Despite their occasional successes, special inspection authorities and requests for additional information are highly limited, so the IAEA has very limited means for confirming or disconfirming its suspicions about undeclared nuclear facilities in states that only have CSAs.

\section{The role of intelligence agencies in identifying undeclared nuclear facilities} \label{theroleofintelligenceagenciesinidentifyingundeclarednuclearfacilities}

It appears that, unofficially, the IAEA identifies suspect locations mainly through voluntary tips from national intelligence agencies (and perhaps also from whistleblowers). These tips from intelligence agencies appear to be irreplaceable for the IAEA’s ability to identify potential undeclared nuclear facilities. These conclusions are supported by several sources of evidence:

\textbf{Expert opinion}: Experts appear to agree that intelligence agencies play an irreplaceable and primary role in detecting undeclared nuclear facilities.\footnote{What do experts at the IAEA say? The IAEA gives few details about its processes for identifying locations that might host undeclared nuclear facilities, although it occasionally refers vaguely to information from "third parties"  \cite{iaea2017}. An IAEA report expresses confidence in the abilities of intelligence agencies to detect undeclared facilities: "it is probable that [intelligence] services would detect the construction by another State of any significant facility which should have been but was not reported to the IAEA under the provisions of the relevant safeguards agreement. The State which discovered the facility would be free to draw the fact to the attention of the Board of Governors"  \cite{iaea1983}.}

\begin{itemize}
	\item Leonard Spector, then-director of the Nuclear Non-Proliferation Project at the Carnegie Endowment, explained, "U.S. intelligence [...] has been the principal, if announced, mechanism for detecting [Non-Proliferation] treaty violations…"  \cite{spector1992}.
	\item Michael O’Hanlon, Director of Research of the Foreign Policy program at the Brookings Institution, writes, "[T]he so-called "Additional Protocol" has created the right for inspectors to go to places where they suspect monkey business, even if those sites are not officially declared by the country in question. This arrangement tends to work only if national intelligence capabilities, and/or whistleblowers, provide information about suspicious activities. But at that point, inspectors can be more effective than in the years before the Additional Protocol concept was developed and legitimated"  \cite{ohanlonetal2020}.
	\item A report from the Australian Safeguards and Non-Proliferation Office asserts, "there is no doubt that national intelligence information will continue to have a vital role in the detection of undeclared nuclear activities"  \cite{carlsonetal2006}.
	\item Across a wide range of relevant publications (those read for this research), there is no mention of experts expressing contrary opinions.

\end{itemize}
\textbf{Case studies}: National intelligence agencies have historically had a primary role in prompting IAEA investigations of undeclared nuclear facilities. Of the 4 states in which the IAEA has investigated what turned out to be undeclared nuclear facilities (Syria, North Korea, Iran, and Iraq\footnote{These states, as well as Libya, South Korea, and Yugoslavia, are the NPT non-nuclear-weapon states that pursued nuclear weapons  \cite{bleek2017}. However, the latter three did not have secret nuclear facilities inspected by the IAEA. (The IAEA  \cite{directorgeneral2004} just inspected Libya’s "previously undeclared" nuclear facilities.)}), there have been 3 states in which the IAEA began to investigate the facilities mainly or entirely because of intelligence agency tips, while there have been no states in which the IAEA began to investigate them mainly or entirely through its own CSA or AP-authorized processes.\footnote{As a limitation of these examples’ scope, they are almost all about states that did not have APs in force.}

\begin{itemize}
	\item In 3 cases (Syria, North Korea, and Iran), the IAEA mainly or entirely learned of the undeclared facilities from intelligence agency tips.
	\begin{itemize}
		\item Western intelligence agencies notified the IAEA of an undeclared nuclear reactor (after Israel bombed it) and three additional suspected facilities in Syria, which the IAEA then investigated  \cite{boardofgovernors2011}\cite{harelandbenn2018}.
		\item The IAEA investigated undeclared nuclear sites in North Korea based on U.S. tips  \cite{fischer1997}.
		\item The IAEA investigated the first two undeclared nuclear facilities in Iran based on tips from intelligence agencies.\footnote{These facilities are commonly believed to have been revealed by an Iranian dissident group, but some analysts compellingly argue that the U.S. intelligence community tipped off the IAEA first  \cite{lewis2006a}.}

		\item Then, Iran notified the IAEA of its other undeclared facility (FFEP) shortly before Western states announced it. Reportedly, "the letter was only sent after the Iranian government discovered the secret plant had been discovered by western intelligence"  \cite{traynorandborger2009}.
		\item The IAEA has also investigated additional undeclared nuclear activities in Iran  \cite{directorgeneral2022}. It does not report having been first to spot any of the relevant locations, and at least some of these investigations appear to have been based on tips from Western intelligence  \cite{dichristopher2018}.
	\end{itemize}
	\item The other case was Iraq, where the IAEA was acting with UNSC-granted authorities well beyond its usual ones (due to Iraq’s defeat in the Gulf War).

\end{itemize}
\textbf{Plausible mechanism}: Intelligence agencies have uniquely strong intelligence gathering capabilities, which could explain their unique contributions.\footnote{Compared to the IAEA or nonprofit analysts, intelligence agencies are presumably much more capable of (e.g., they have much more experience with) spying, intercepting electric signals, and monitoring relevant supply chains. They are also more capable of satellite imagery analysis, because they have e.g. higher-resolution photos (and this advantage was even more extreme historically, before the rise of commercial satellite imagery  \cite{zegart2022}).}

\section{Details on track records of nuclear arms control M\&V} \label{detailsontrackrecordsofnucleararmscontrolmv}

\textbf{No state has attempted to divert a significant quantity of nuclear material from a facility under CSA safeguards.}

\begin{itemize}
	\item We can draw this conclusion on the grounds that there is no mention of such diversion across a varied and fairly comprehensive range of materials on relevant cases.\footnote{These include all sources cited in this report.}

	\item The IAEA recently found that Iran has some undeclared nuclear material  \cite{directorgeneral2022}. However, there appears to be no public evidence that it came from declared facilities, nor that the amounts of nuclear material involved were significant for proliferation. Additionally, the IAEA’s  \cite{iaea2022c} subsequent annual safeguards statement concluded that Iran’s "declared nuclear material remained in peaceful activities."
	\item There have been a few cases (e.g., Yugoslavia, South Korea) of tiny amounts of nuclear materials (reportedly) having been diverted from facilities that were under CSA safeguards. However, the relevant quantities were far below the "significant quantities" of diversion that CSA safeguards are designed to notice  \cite{koch1997}\cite{nti2015d}.
	\item Iraq and Yugoslavia planned to divert nuclear materials that were under CSA safeguards, but they never executed these plans  \cite{potteretal2000}\cite{albright2002}\cite{nti2015a}.\footnote{After international backlash to its invasion of Kuwait, Iraq’s government planned to divert uranium from a safeguarded reactor, but this plan was canceled after other equipment—which was necessary for weaponizing the uranium—was accidentally destroyed by the U.S.-led coalition’s bombing. Yugoslavia’s nuclear weapons program planned on getting nuclear materials from its civil nuclear infrastructure, but the program was canceled before they reached that step.}
\end{itemize}
\textbf{The U.S. likely discovered Russia’s violation of the INF Treaty before it offered Russia a substantial strategic advantage, and the U.S. plausibly had reliable monitoring throughout.}

\begin{itemize}
	\item According to a statement from the U.S. Director of National Intelligence (consistent with a State Department report), on one hand, the U.S. took about 5 years after Russia began the missile’s development to raise accusations  \cite{coats2018}\cite{state2019}.
	\item On the other hand,
	\begin{itemize}
		\item the statement suggests the missile’s testing program was not completed for another 2 years;
		\item the statement suggests the missile may have been impossible to identify as a violation until later stages of this testing program; and
		\item U.S. officials reportedly stated—5 years after the initial U.S. allegations—that Russia had still deployed fewer than 100 of the violating missiles (less than one fifteenth the number of intermediate-range missiles the U.S.S.R. had before the treaty)  \cite{gordon2018}.

	\end{itemize}
\end{itemize}
\textbf{Beside the INF Treaty incident, the U.S. has mostly confirmed the U.S.S.R./Russia’s compliance across the INF Treaty, START I, and New START.}

\begin{itemize}
	\item The U.S. Director of National Intelligence spoke warmly of initial compliance on the INF Treaty: "Together, we eliminated over 2,600 prohibited missiles"  \cite{coats2018}.
	\item In 2001, the U.S. State Department announced, "The [START I’s] final ceilings came into effect today, and they have been met"  \cite{powell2001}.
	\item Agreeing with the above, NTI’s page on the START I mentions no accusations of serious non-compliance  \cite{nti2011}.
	\item The U.S. State Department states, "Although the United States has raised implementation-related questions and concerns with the Russian Federation through diplomatic channels and in the context of the BCC, the United States has determined annually since the treaty’s entry into force, across multiple administrations, the Russian Federation’s compliance with its treaty obligations"  \cite{state2023}.

\end{itemize}
\textbf{Russia has made several relatively minor or tenuous allegations of noncompliance with the INF Treaty, START I, and New START.}

\begin{itemize}
	\item In 2001, in the context of START I, Russia disputed whether the U.S. had destroyed enough of the stages of one type of missile  \cite{nti2011}. Russia also alleges that several non-secret U.S. missile-related activities violate the INF Treaty. Some of Russia’s allegations—like that U.S. drones count as cruise missiles—appear tenuous  \cite{woolf2019}.
	\item Additionally, the U.S. and Russia suspended New START inspections with the COVID-19 pandemic (unlike the IAEA, which continued its inspections)  \cite{iaea2020a}\cite{bugos2022}. Then, during the Russian invasion of Ukraine, Russia suspended its participation in New START altogether. However, whether or not this is noncompliance, it is not a case of subtle circumvention of the M\&V system.

\end{itemize}
\textbf{In U.S.-U.S.S.R./Russia nuclear arms control treaties other than the above, parties made non-compliance accusations, but they did not raise accusations of massive non-compliance for ratified treaties.}

\begin{itemize}
	\item The U.S. and the U.S.S.R. accused each other of developing radar systems in violation of the Anti-Ballistic Missile Treaty. Both parties also claimed other less specific or clear violations  \cite{reagan1987}\cite{federationofamericanscientists1988}.
	\item Reagan accused the Soviets of having "violated the [SALT I] prohibition on the use of former ICBM facilities"  \cite{reagan1987}.
	\item Reagan accused the Soviets of violating core provisions (including strategic arms limits) of the SALT II Treaty, but he also noted this treaty was never ratified and would have expired if it had been  \cite{reagan1987}.

\end{itemize}
\textbf{There have been no significant, known violations of nuclear test ban treaties, except possibly in the Vela incident.}

\begin{itemize}
	\item A wide range of sources\footnote{These include the sources cited in this paper.} make no mention of serious accusations of violations other than the \hyperref[43nucleartestbanmvsystemstrackrecords]{Vela incident}.

	\item An article published on the IAEA bulletin on the 10th anniversary of the Partial Test Ban Treaty states, "The record of compliance with the PTB is generally considered to be good. There has so far been no complaint of a significant breach by any party"  \cite{delcoigneetal1973}.
	\item Radioactive carbon in the atmosphere peaked right around the signing of the Partial Test Ban Treaty  \cite{universiteitutrechtnd}.
	\item Additionally, the CTBT was signed in 1996, and only a few non-signatories are known to have conducted nuclear tests since then  \cite{ourworldindata2022}.
	\item Reagan accused the Soviets of violating the Threshold Test Ban Treaty, but this was before either party had ratified the treaty, and concerns were later resolved  \cite{reagan1987}\cite{lewis2018}.

\end{itemize}
\section{How verification of AI chip accounts could be implemented} \label{howverificationofaichipaccountscouldbeimplemented}

\subsection{Introduction} \label{introduction}

For verifying certain international agreements on AI, \textbf{it would be helpful (though not sufficient) to be able to verify the quantity, location, and integrity}\footnote{Here, "integrity" refers to having whichever hardware features they are supposed to have (according to some agreement on AI), i.e. having been designed appropriately and not tampered with.}\textbf{ of all cutting-edge, AI-specialized chips}\footnote{Precisely which types of chips, equipment, or data centers would need to be monitored is an open question.}, at least well enough to detect very large numbers of these chips that have been tampered with or are kept secret. Shavit  \cite{shavit2023} describes one way AI chip accounts could fit into a broader verification regime, and he briefly discusses how these accounts could be verified. \textbf{This appendix aims to provide a detailed description of how AI chip accounts could be implemented with multiple layers of verification}, mainly using methods with close analogues in nuclear arms control verification.

This potential implementation is not a policy suggestion. Instead, it is described to show that reliable verification of AI chip accounts \emph{could} be done mainly with methods that have \hyperref[41nptmvsystemstrackrecords]{successful} historical analogues. We focus on these methods here because we have more information about their political feasibility, but that is insufficient to make an overall recommendation.

To avoid repetition, this appendix assumes basic familiarity with nuclear M\&V methods, which are described in \hyperref[2nuclearmvmethods]{a previous section}.

\subsection{Background treaty obligations and verification goals} \label{backgroundtreatyobligationsandverificationgoals}

This appendix is about the verification of a hypothetical international agreement that obliges state parties to the following:

\begin{itemize}
	\item \textbf{Chip features: }Ensure that cutting-edge, AI-specialized chips ("AI chips") in the state’s territory continually have certain design (and/or assembly) features, including a \hyperref[223uniqueidentifiers]{unique identifier}; and
	\item \textbf{AI chip accounting:} Notify and regularly update some international organization of the existence, locations, technical specifications, and \hyperref[223uniqueidentifiers]{unique identifiers} of cutting-edge AI chips and of the machines that make them.
	\item \textbf{Location restrictions on data-center-quality AI chips:} Keep data-center-quality AI chips in production facilities, data centers, storage facilities, elimination facilities\footnote{Here, "elimination facilities" refers to facilities that destroy chips that will no longer be used (e.g. for recycling their metals).}, or in (time-limited) transit between these locations.\footnote{This limits the scope of inspections needed, though additional measures may be needed to let individuals have commodity chips without this being a loophole.}
\end{itemize}
This appendix assumes that the main types of violations this system is aiming to detect are: the possession of many cutting-edge AI chips at undeclared locations, and the possession of many cutting-edge AI chips that lack required design features. The problem of detecting these violations can be broken down into the following two subproblems:

\begin{enumerate}
	\item Verify that there are not many\footnote{Technically, the term "many" should refer to a lower number in the subproblems than in the overall problem, so that the margin of error that is accepted when addressing the subproblems does not add up to an overall margin of error that is too high.} cutting-edge AI chips being used at undeclared locations; and

	\item Verify that, of the cutting-edge AI chips at reported locations, not many lack required design features.

\end{enumerate}
\subsection{Detecting efforts to get cutting-edge AI chips to undeclared locations} \label{detectingeffortstogetcuttingedgeaichipstoundeclaredlocations}

To verify that there are not many cutting-edge AI chips being used at undeclared locations, one approach is: verify that these chips could not have reached undeclared locations, on the grounds that \textbf{any attempt at secretly producing cutting-edge AI chips or removing them from their declared locations would have been detected}. This section describes how that detection could be done, for each potential method of secretly producing or moving these chips.

First, AI chips could be secretly produced by accurately reporting the existence of chip-making machines but \textbf{under-reporting production levels}. To detect this, one could:

\begin{itemize}
	\item Have in-line instrumentation be installed on chip-making machines (or power systems) to monitor production levels;
	\item Monitor relevant chips fabrication facilities’ ("fabs’") \hyperref[234procurementmonitoring]{procurement activities}, looking for undeclared purchases of chip manufacturing materials; and
	\item Establish perimeter portal continuous monitoring at relevant fabs, looking for undeclared shipments of chips out of the fabs.

\end{itemize}
Alternatively, AI chips could be secretly produced by \textbf{undeclared chip-making machines} at declared fabs. To detect this, one could:

\begin{itemize}
	\item Carry out inspections at relevant fabs, implemented like the "undeclared item inspections" described in \hyperref[detectingeffortstohostcuttingedgeaichipsatundeclaredlocations]{the next section} (except in this case, inspectors would be looking for undeclared chip-making machines, rather than undeclared chips);
	\item Implement \hyperref[224designinformationverification]{design information verification} at relevant fabs, to detect secret rooms;
	\item Monitor relevant fabs’ \hyperref[234procurementmonitoring]{procurement activities}, looking for undeclared purchases of chip manufacturing materials or chip-making machines (potentially along with video surveillance and \hyperref[225perimeterportalcontinuousmonitoring]{perimeter monitoring}, for especially centralized suppliers); and
	\item Establish \hyperref[225perimeterportalcontinuousmonitoring]{perimeter portal continuous monitoring} at relevant fabs, looking for undeclared shipments of chips out of the fabs or of chip-making machines or machine components into the fabs.

\end{itemize}
Thirdly, AI chips could theoretically be secretly produced at \textbf{undeclared fabs}. To detect this, one could:

\begin{itemize}
	\item Use \hyperref[231nationaltechnicalmeans]{national technical means} to look for signs of secret fab construction;
	\item \hyperref[234procurementmonitoring]{Monitor the sales} of fab suppliers, looking for undeclared purchases of chip manufacturing materials or chip-making machines; and
	\item Use \hyperref[236challengeinspectionsandrequestsforadditionalinformation]{challenge inspections} to resolve suspicions about particular locations.

\end{itemize}
Instead of secretly producing cutting-edge AI chips, an adversary could seek to \textbf{secretly divert cutting-edge AI chips from their reported locations at data centers and storage facilities}. To detect this, one could:

\begin{itemize}
	\item Carry out "diversion detection inspections"\footnote{This and the other types of inspections in quotes are not terms from nuclear arms control.} at declared data centers and storage facilities (as described by Shavit  \cite{shavit2023}, these could consist of inspectors specifying a random sample of unique identifiers that are reportedly at some facility, facility operators providing access to the corresponding AI chips, and then inspectors checking that the reported chip is present\footnote{Additionally, to ensure inspectors have inspection access to enough AI chips (rather than storage facilities always having the option to say they recently shipped out chips inspectors are asking to see), data centers and storage facilities could be required to hold AI chips for some minimum period between receiving and shipping them (similarly to how the IAEA uses "mailbox declarations").}); and

	\item Use video surveillance at declared data centers and storage facilities, as additional measures to detect diversion.

\end{itemize}
Alternatively, an adversary could attempt to \textbf{secretly divert AI chips from their reported transit paths}. To detect this, one could:

\begin{itemize}
	\item Use video surveillance throughout transit; and
	\item Carry out "diversion detection inspections" (as \hyperref[detectingeffortstogetcuttingedgeaichipstoundeclaredlocations]{above}) at declared data centers and storage facilities, which could detect a diversion during recent transit.

\end{itemize}
Lastly, an adversary could attempt to \textbf{secretly divert AI chips or chip-making machines from an elimination facility}, e.g. by falsely reporting that chips have been melted. To detect this, one could use "elimination inspections," designed with a few measures to detect this violation:

\begin{itemize}
	\item Inspector presence over specified elimination procedures; and
	\item \hyperref[221onsitemeasurementmethods]{Unattended surveillance equipment}.

\end{itemize}
Alternatively, a mandated storage period before elimination could help ensure that AI chips would be unusable or obsolete by the time they are eliminated.

\subsection{Detecting efforts to host cutting-edge AI chips at undeclared locations} \label{detectingeffortstohostcuttingedgeaichipsatundeclaredlocations}

So far, we have examined one approach to verifying that there are not many cutting-edge AI chips being used at undeclared locations: detecting attempts to get cutting-edge AI chips to these locations. Another possible approach is \textbf{detecting cutting-edge AI chips after they have reached undeclared locations, when they are being stored or used}. This section describes how that could be implemented, for each potential method of secretly hosting operational, cutting-edge AI chips.

Cutting-edge AI chips could be secretly stored at facilities that are not declared to be data centers, which would effectively be \textbf{undeclared data centers}. To detect these data centers, one could:

\begin{itemize}
	\item Use \hyperref[231nationaltechnicalmeans]{national technical means} (e.g. satellite images);
	\item \hyperref[234procurementmonitoring]{Monitor procurement} of equipment that is used for building and operating data centers;
	\item Carry out \hyperref[236challengeinspectionsandrequestsforadditionalinformation]{challenge inspections} to resolve suspicions about particular locations; and
	\item Specifically for verifying that AI chip storage facilities are not being used as data centers, \hyperref[224designinformationverification]{design information verification} and "undeclared item inspections" (described \hyperref[detectingeffortstohostcuttingedgeaichipsatundeclaredlocations]{below}).

\end{itemize}
Alternatively, cutting-edge AI chips could be stored at \textbf{declared data centers}, without being reported as being there. These could be detected with "undeclared item inspections"\footnote{This refers to the inspector activities described below, except the design information verification.} along with \hyperref[224designinformationverification]{design information verification}, as follows.

\begin{itemize}
	\item AI chips might be "hidden in plain sight" in \textbf{declared server rooms}.\footnote{Room functions could be declared as parts of design information verification.} To detect this, one could use "undeclared item inspections," with:

	\begin{itemize}
		\item Counting of chips in server rooms; and
		\item Examination of \hyperref[223uniqueidentifiers]{unique identifiers} of selected chips in server rooms.
	\end{itemize}
	\item AI chips might be hidden in \textbf{areas that reportedly do not contain chips}, at declared data centers. To detect this, one could use "undeclared item inspections," with:
	\begin{itemize}
		\item Visual observation and equipment that can detect the presence of chips; and
		\item Surveillance of declared non-server areas with unattended equipment, e.g. using video cameras to verify that an area is not used for chip storage.
	\end{itemize}
	\item AI chips might be hidden in \textbf{secret rooms}, at declared data centers. To detect this, one could use \hyperref[224designinformationverification]{design information verification}, with:
	\begin{itemize}
		\item Design information verification inspections;
		\item Unattended on-site equipment, e.g. (if needed, rudimentary) video and audio surveillance equipment to detect construction activities; and
		\item Satellites, to detect undeclared construction.
	\end{itemize}
	\item AI chips that have been placed in a data center might be hidden by \textbf{being moved}, during an inspection, to areas that will not undergo (further) inspection. To detect this, one could include the following procedures in "undeclared item inspections," done by \hyperref[22mvmethodsatdeclaredfacilities]{inspectors monitoring the exits} of rooms and facilities:
	\begin{itemize}
		\item Inspection of items leaving not-yet-inspected rooms; and
		\item Inspection of items leaving the data center, as well as on-site video surveillance.

	\end{itemize}
\end{itemize}
Zooming out, this approach could be combined with the one discussed in \hyperref[detectingeffortstogetcuttingedgeaichipstoundeclaredlocations]{the previous section} for redundancy. Next, we turn to the other subproblem of verification.

\subsection{Detecting efforts to tamper with the design features of cutting-edge AI chips at known locations} \label{detectingeffortstotamperwiththedesignfeaturesofcuttingedgeaichipsatknownlocations}

To detect efforts to secretly tamper with large numbers of AI chips at known locations, one could use the following:\footnote{In addition to inspecting chips shortly after their production, it might also be feasible to directly inspect chip-making machines, to verify that they are built to introduce the desired devices from the start. However, that might be significantly more technically challenging and operationally disruptive, while potentially also raising stronger information security concerns.}

\begin{itemize}
	\item \textbf{"Tampering detection inspections"} at data centers and storage facilities (including soon after chip production) could help verify that chips have not been tampered with.
	\begin{itemize}
		\item These could consist of inspectors specifying a random sample of chips (identified by their unique identifiers), data center operators providing access to them, and inspectors then verifying that the chips have not been tampered with.\footnote{It may be most efficient for these to be done concurrently with other types of inspections, as is suggested by New START’s consolidation of inspection types.}

	\end{itemize}
	\item \textbf{\hyperref[222containmentandsurveillance]{Video surveillance}} at data centers, storage facilities, and in transit could directly detect tampering activities.

\end{itemize}
\subsection{Additional measures} \label{additionalmeasures}

The three approaches discussed above could each be boosted by \hyperref[232humansources]{human sources} of information, who could be accessed by broad interview authority and collection of tips.

For efficiency and preventing cover-ups, inspections should often be short-notice, randomly timed, and with heavy reliance on random sampling within the inspection (when that is sufficient for a high probability of detecting violations\footnote{This is the case when a violation would involve a large number of non-compliant items and a not-vastly-larger number of compliant items; then, random (independent) sampling tends to quickly find non-compliant items (by the central limit theorem).}).

\subsection{Overview of the hypothetical implementation described above (not a policy suggestion)} \label{overviewofthehypotheticalimplementationdescribedabovenotapolicysuggestion}

\textbf{At certain production and storage facilities upstream in the supply chain:}

\begin{itemize}
	\item \hyperref[21accountingandmandatoryselfreporting]{Accountancy reporting} (of cutting-edge chip-making machines)
	\item \hyperref[234procurementmonitoring]{Procurement monitoring}
	\item \hyperref[222containmentandsurveillance]{Video cameras} (perhaps) (focused on highly centralized suppliers)
	\item \hyperref[225perimeterportalcontinuousmonitoring]{Perimeter portal continuous monitoring} (perhaps) (focused on highly centralized suppliers)

\end{itemize}
\textbf{At cutting-edge fabs:}

\begin{itemize}
	\item \hyperref[21accountingandmandatoryselfreporting]{Accountancy reporting} (of cutting-edge AI chips and chip-making machines)
	\item \hyperref[detectingeffortstohostcuttingedgeaichipsatundeclaredlocations]{Undeclared item inspections} (focused on chip-making machines)
	\item \hyperref[detectingeffortstotamperwiththedesignfeaturesofcuttingedgeaichipsatknownlocations]{Tampering} and \hyperref[detectingeffortstogetcuttingedgeaichipstoundeclaredlocations]{diversion detection inspections}
	\item \hyperref[224designinformationverification]{Design information verification}
	\item \hyperref[225perimeterportalcontinuousmonitoring]{Perimeter portal continuous monitoring} (focused on cutting-edge AI chips, chip-making machines, and construction items)
	\item In-line instrumentation

\end{itemize}
\textbf{At cutting-edge data centers and cutting-edge AI chip storage facilities:}

\begin{itemize}
	\item \hyperref[21accountingandmandatoryselfreporting]{Accountancy reporting} (of cutting-edge AI chips)
	\item \hyperref[detectingeffortstohostcuttingedgeaichipsatundeclaredlocations]{Undeclared item inspections}
	\item \hyperref[detectingeffortstotamperwiththedesignfeaturesofcuttingedgeaichipsatknownlocations]{Tampering} and \hyperref[detectingeffortstogetcuttingedgeaichipstoundeclaredlocations]{diversion detection inspections}
	\item \hyperref[222containmentandsurveillance]{Video cameras}
	\item \hyperref[224designinformationverification]{Design information verification}

\end{itemize}
\textbf{At certain elimination facilities:}

\begin{itemize}
	\item \hyperref[21accountingandmandatoryselfreporting]{Accountancy reporting} (of AI chips and chip-making machines)
	\item \hyperref[detectingeffortstogetcuttingedgeaichipstoundeclaredlocations]{Elimination inspections} (focused on AI chips and chip-making machines)
	\item Unattended surveillance equipment
	\item Storage period before elimination (potential alternative)

\end{itemize}
\textbf{In transit:}

\begin{itemize}
	\item \hyperref[222containmentandsurveillance]{Video cameras} (for cutting-edge AI chips and chip-making machines)
	\item GPS devices and security for vehicles

\end{itemize}
\textbf{At arbitrary locations:}

\begin{itemize}
	\item Location restrictions on data-center-quality, cutting-edge AI chips
	\item \hyperref[231nationaltechnicalmeans]{National technical means}
	\item \hyperref[236challengeinspectionsandrequestsforadditionalinformation]{Challenge inspections}
	\item \hyperref[232humansources]{Human sources}

\end{itemize}
\subsection{Assessment} \label{assessment}

There are good reasons to tentatively consider the verification system described above \textbf{highly reliable}:\footnote{One issue that has been glossed over is attribution; enforcement of treaties with more than two parties presumably requires knowing not just whether a violation occurred but also who committed it. Still, with the addition of security requirements across the supply chain (to make discreet theft more difficult), the system as described above seems plausibly adequate for attribution. This is because violations would be tied to a specific facility (or vehicle), facility (or vehicle) owner, and host state (and potentially to a specific international heist), and these could be investigated further if a violation were detected.}

\begin{itemize}
	\item It has at least \textbf{5 layers of defense}\footnote{From counting the detection measures discussed above, we can see that the system has at least two layers of defense for detecting each potential method to get AI chips to undeclared locations, as well as two or more additional layers for detecting each potential method to store and use AI chips at undeclared locations. This reasoning assumes that there are no potential methods to possess and use many AI chips at undeclared locations other than the methods that are considered and countered above; we can assume this because:

	\begin{itemize}
		\item Chips at undeclared locations can have either been produced secretly or not, in which case (since location reporting is required) they must have at some point been moved from a declared location to an undeclared location.
		\item If all sites containing chip-making machines, all the chip-making machines at these sites, and their chip production levels are all truthfully reported on, then there must be no secret production of chips.
		\item If chips whose production is truthfully declared are not diverted from their production facilities or any other places where they are permitted to be, then they are not diverted.

	\end{itemize}
    Finally, the system has a further layer: human sources.} for detecting any serious attempt to possess and use many cutting-edge AI chips at undeclared locations, and it has \textbf{3 layers of defense}\footnote{These are: tampering detection inspections, video surveillance, and human sources} for detecting tampering with chips at declared locations.

	\item \textbf{Some of these layers appear highly reliable}\footnote{These layers appear highly reliable (i.e. appear capable of achieving an at at least a 90\% detection probability for each attempted violation):

	\begin{itemize}
		\item For verification that chips do not reach undeclared locations, the combination of the following: in-line instrumentation, the combined methods for detecting undeclared chip-making machines, diversion detection inspections, and a required waiting period before elimination of old AI chips
		\item For verification of non-tampering, the above (to verify that there are not many AI chips that have been tampered with and are at secret locations, since not many AI chips are at secret locations) together with tampering detection inspections	\end{itemize}}, and all appear to have a significant chance of detecting violations.
	\item \textbf{Analogous methods have worked very well historically} (as discussed in the section on track records).

\end{itemize}
Overall, this shows that \textbf{methods that have been widely used for nuclear arms control verification}\footnote{That is, for (former) U.S.-U.S.S.R./Russia nuclear arms control verification and for the IAEA’s verification of the Non-Proliferation Treaty}\textbf{ can be adapted to create a reliable system for verifying accounts of AI chips}.

\section{Back-of-the-envelope calculations of direct inspection costs} \label{backoftheenvelopecalculationsofdirectinspectioncosts}

\textbf{Across three metrics}—the number of inspections conducted, the ratio of number of items examined in inspections to gross world product, and the proportional loss to gross world product from interruptions to facility (i.e. data center) operations—\textbf{back-of-the-envelope calculations suggest that the direct costs of inspecting cutting-edge AI chips} to verify the compliance of highly compute-intensive AI development would be \textbf{lower than or roughly similar to the corresponding costs in IAEA verification} (i.e. less than 10x larger).

The last two metrics incorporate gross world product to show that, although speculative large increases in global AI chip production would increase the number of chips that needed to be examined and potentially interrupted to verify compliance, such increases in chip production would involve enough economic growth to easily keep pace with this need.\footnote{In other words, the argument here is that, even if AI chip production grew exponentially, this would not involve much proportionally larger costs than IAEA inspections involve. We may assume proportional costs are more relevant to stakeholders than absolute costs.}

\textbf{Number of inspections conducted annually:}

\begin{itemize}
	\item In 2019 and 2020, the IAEA conducted ~2,900 in-field inspections  \cite{iaea2020b}\cite{iaea2021a}.
	\item Various estimates suggest there are currently ~100-1,000 large data centers, and an analyst at market research group Synergy Research Group claims this figure has doubled over the past ~5 years  \cite{woodie2019}\cite{haranas2021}\cite{strohmaieretal2022}\footnote{The dataset of Strohmaier \emph{et al}. shows that the 500th highest-performing data center has ~0.16\% the throughput of the highest-performing one and ~14\% the throughput of the 50th-highest.}. The number of data centers with cutting-edge AI chips is presumably significantly lower.

	\item The number of chip fabrication facilitates ("fabs") producing (near-)cutting-edge chips appears relatively small.\footnote{Prominent semiconductor manufacturing companies such as TSMC  \cite{tsmc2021}, Samsung  \cite{samsunggroup2022}, Intel  \cite{intelcorporation2022}, and SMIC  \cite{smic2020} each report under 20 modern semiconductor manufacturing facilities.}

	\item This leaves substantial room for the number of large, AI-specialized data centers and fabs to grow while the number of inspections needed to inspect them stays roughly similar to the IAEA’s number of inspections.
	\begin{itemize}
		\item Even if demand for data centers and fabs hosting a significant number of cutting-edge, AI-specialized chips exploded upwards, their numbers could be kept manageable, e.g. through a large but limited number of authorizations of cutting-edge, AI-specialized data centers.

	\end{itemize}
\end{itemize}
\textbf{Ratio of the number of items examined annually to gross world product:}

\begin{itemize}
	\item In 2019 and 2020, the IAEA verified that ~25,000 seals had not been tampered with  \cite{iaea2020b}\cite{iaea2021a}. Additionally, IAEA inspectors conduct many other measurements, such as full physical inventories at most of 700 facilities, which amounts to annual examinations of ~140,000 items if we assume 80\% of facilities are inspected and 250 items\footnote{This seems conservative; an IAEA report explains that, "The inventory in the facility [...] may be upwards of a hundred tonnes of material in a variety of forms such as pure powder, assemblies, rods, unsintered pellets, sintered pellets, reject materials awaiting recycle, and scrap material in a variety of inhomogeneous forms"  \cite{iaea1980}.} are examined in the average inventory  \cite{henriques2016}.

	\item Assuming that 10k to 100k AI chips\footnote{The PaLM paper used ~6,000 chips over ~2 months  \cite{chowdheryandnarang2022}.} used for a year were required for a violation and assuming (very conservatively relative to current AI chip production  \cite{rahmat2022}\cite{morgan2022}) that the number of chips that could be used for a violation were 2.5 billion, then ~58,000 to ~580,000 chips would have to be examined annually for a 90\% probability of detecting serious violations.\footnote{A general formula, derivable by assuming that random sampling of items is done independently and with replacement (which is a slightly conservative approximation of sampling without replacement), is that the number of chips that need to be examined is $F \cdot \mbox{log}(1-p)/\mbox{log}(1-M/N)$, where p is the probability of detecting a violation, M is the number of chips needed for a serious violation, N is the total number of chips one is sampling from, and F is the annual frequency at which examinations need to occur. This is a slight variant of the equation proposed by Shavit  \cite{shavit2023}; his equation is approximately equivalent to this one, given the parameter choices in this report (since, then, $e^{-fT}$ is negligible, and $M/N \approx H/(aTC)$).} The GWP involved would not be less than it currently is for IAEA verification, so the ratio (of items examined to GWP) would be less than 10x worse (and plausibly better) than the ratio involved in IAEA inspections.

	\begin{itemize}
		\item Hypothetically, what if the number of available AI chips grew beyond that already high level? 2x growth in AI chip production would cause a ~2x increase in the number of chips that need to be examined to maintain a 90\% detection probability.\footnote{This can be computed from the formula in the previous footnote (for a wide range of multipliers, from 2x to e.g. 1,000,000x).} At the same time, such an increase in AI chip production from such a high level would require ~2x growth in GWP\footnote{From a starting point of 2.5 billion AI chips, each costing (let us conservatively assume) \$1,000 each, spending on AI chips would already be \$2.5 trillion, which is estimated to be approximately the current global spending on R\&D  \cite{sargent2022}. Assuming that the ratio of R\&D spending to GWP stayed constant and that the vast majority of AI chips were used for R\&D (i.e. AI training runs), a ~2x increase to the number of chips would thus require a ~2x increase to GWP. Even conservatively assuming that AI chip spending as a fraction of GWP could rise to 20\%, the resulting increase in the ratio of items examined to GWP would be bounded by a factor of ~8, and that is conservatively assuming that \$2.5 trillion were spent on AI chips with no GWP growth.}, so the ratio would not become worse.
	\end{itemize}
\end{itemize}
\textbf{Proportional loss of gross world product from interruptions to operations:}

\begin{itemize}
	\item Nuclear energy sales make up ~0.4\% of gross world product  \cite{gillespie2022}. Assuming the average nuclear energy facility loses 2 days of production per year to IAEA safeguards inspections\footnote{An IAEA publication, though it does not report average lengths of time for which inspections disrupt nuclear facilities’ operations, writes, "A physical inventory verification at a large facility can be so complex and time-consuming that it might take up to 10 inspectors 7 to 14 days to complete [...] physical inventory verifications are conducted once a year in most of the 700 facilities that are under IAEA safeguards worldwide"  \cite{henriques2016}. Other IAEA publications  \cite{iaea1980}\cite{iaea2015} clarify that inventories (tend to) involve a shut-down.}, it follows that the interruptions from these inspections cost ~0.002\% of gross world product.

	\item Assuming 10,000 AI chips for a year out of anywhere from 600,000 to 600 billion (or even more) available AI chips were required for a violation, making the simplifying (and very conservative) assumption that AI chips made up all of gross world product, and assuming that examination of any one chip knocked 100 chips out of use for one day per year, these inspections would cost ~0.006\% of gross world product\footnote{This follows from the formula in the previous footnote. In the context of the assumptions made above, that formula implies ~0.023\% of the sampled-from chips are selected for examination (across the whole range of N), and 0.023\%*100/365 ~= 0.006\%.}—just a factor of ~3 over the above estimate about IAEA inspections’ costs.

	\begin{itemize}
		\item This estimate assumes that fab production levels can be verified without interrupting fab production (e.g. by monitoring chip-making machines’ inputs and outputs) and are therefore relatively small. Otherwise, costs could be much higher.

	\end{itemize}
\end{itemize}
These are just back-of-the-envelope calculations made with extremely rough approximations, so they should not be treated as robust predictions.

\end{document}